\newif\ifdoCONF
\doCONFtrue 
\doCONFfalse 
 
\documentclass[PAPER, atlasdraft=false, REPORT=false, UKenglish, cernpreprint=true, texmf, orcidlogo]{atlasdoc}
 
\usepackage[subfig,subcaption,abbrevref=false]{atlaspackage}
\usepackage{atlasbiblatex}
 
\usepackage[hepparticle,hepprocess]{atlasphysics}

\newif\ifshort
\shorttrue   
 
\graphicspath{{logos/}{figures/}}
 
\usepackage{ANA-HDBS-2022-02-PAPER-defs}
\usepackage{multirow}

\AtlasTitle{Search for pair production of boosted Higgs bosons via vector-boson fusion in the $b\bar{b}b\bar{b}$ final state using $pp$ collisions at $\sqrt{s} = \SI{13}{\ensuremath{\text{TeV}}}$ with the ATLAS detector}

\AtlasAbstract{
A search for Higgs boson pair production via vector-boson fusion is performed in the Lorentz-boosted regime, where a Higgs boson candidate is reconstructed as a single large-radius jet, using 140 fb$^{-1}$ of proton--proton collision data at $\sqrt{s} = 13$ TeV recorded by the ATLAS detector at the Large Hadron Collider.
Only Higgs boson decays into bottom quark pairs are considered.
The search is particularly sensitive to the quartic coupling between two vector bosons and two Higgs bosons relative to its Standard Model prediction, \ensuremath{\kappa_{2V}}.
This study constrains \ensuremath{\kappa_{2V}} to $0.55 < \kappa_{2V} < 1.49$ at the \SI{95}{\%} confidence level.
The value $\kappa_{2V} = 0$ is excluded with a significance of $3.8$ standard deviations with other Higgs boson couplings fixed to their Standard Model values.
A search for new heavy spin-0 resonances that would mediate Higgs boson pair production via vector-boson fusion is carried out in the mass range of \SIrange{1}{5}{\ensuremath{\text{T\text{e\kern-0.1em V}}}} for the first time under several model and decay-width assumptions.
No significant deviation from the Standard Model hypothesis is observed and exclusion limits at the 95\% confidence level are derived.
}

\AtlasRefCode{HDBS-2022-02}
 
\PreprintIdNumber{CERN-EP-2024-092}

\AtlasJournalRef{Phys. Lett. B 858 (2024) 139007}
\AtlasDOI{DOI:10.1016/j.physletb.2024.139007}

\hypersetup{pdftitle={ATLAS document},pdfauthor={The ATLAS Collaboration}}
 
\begin{document}
\sisetup{range-phrase=--}
 
\maketitle

\section{Introduction}
The discovery of a \SI{125}{\GeV} Higgs boson (\PH)~\cite{HIGG-2012-27,CMS-HIG-12-028} by the ATLAS~\cite{PERF-2007-01} and CMS~\cite{CMS-TDR-08-001} Collaborations at the Large Hadron Collider (LHC)~\cite{Evans:2008zzb} has led to an extensive research programme aimed at measuring its properties, including spin and parity~\cite{HIGG-2013-17-witherratum,CMS-HIG-14-018,HIGG-2019-01,CMS-HIG-12-041,HIGG-2018-30,HIGG-2018-32,HIGG-2019-10}, natural width~\cite{CMS-HIG-14-018,HIGG-2014-10,CMS-HIG-14-032,CMS-HIG-14-036}, and couplings with other elementary particles~\cite{HIGG-2015-07,HIGG-2021-23,CMS-HIG-22-001}.
All measurements to date are consistent with the predictions from the Standard Model~(SM)~\cite{Englert:1964et,Higgs:1964ia,Higgs:1964pj,Higgs:1966ev,Guralnik:1964eu,Kibble:1967sv}.
However, certain properties, such as quartic couplings ($g_{HHVV}$) with vector bosons ($V = \PW, \PZ$) and the trilinear self-coupling of the Higgs boson ($\lambda_{HHH}$), remain unmeasured.
In the SM, the former are related to the $HVV$ couplings through the relation $g_{HHVV}^\textrm{SM} = g_{HVV}^\textrm{SM}/\nu$, and the latter is predicted to be $\lambda_{HHH}^\textrm{SM} = m_{\PH}^2/2\nu^2$, where $m_{\PH}$ is the Higgs boson mass and $\nu$ is the vacuum expectation value of the Higgs field.
Measuring if these relations are consistent with observed data will fundamentally deepen our understanding of the Higgs mechanism.
 
\begin{figure}[htbp]
\centering
\subfloat[]{\includegraphics[width=0.45\textwidth]{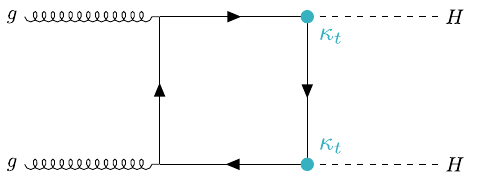}\label{fig:feynman_a}}
\subfloat[]{\includegraphics[width=0.45\textwidth]{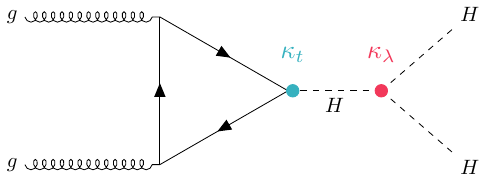}\label{fig:feynman_b}} \\
\subfloat[]{\includegraphics[width=0.3\textwidth]{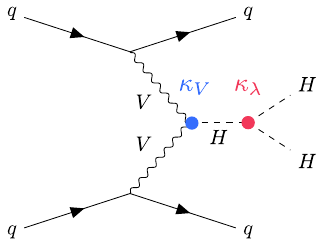}\label{fig:feynman_c}}
\subfloat[]{\includegraphics[width=0.3\textwidth]{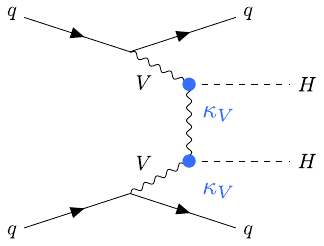}\label{fig:feynman_d}}
\subfloat[]{\includegraphics[width=0.3\textwidth]{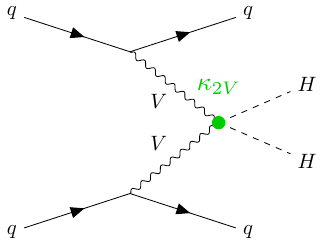}\label{fig:feynman_e}} \\
\subfloat[]{\includegraphics[width=0.33\textwidth]{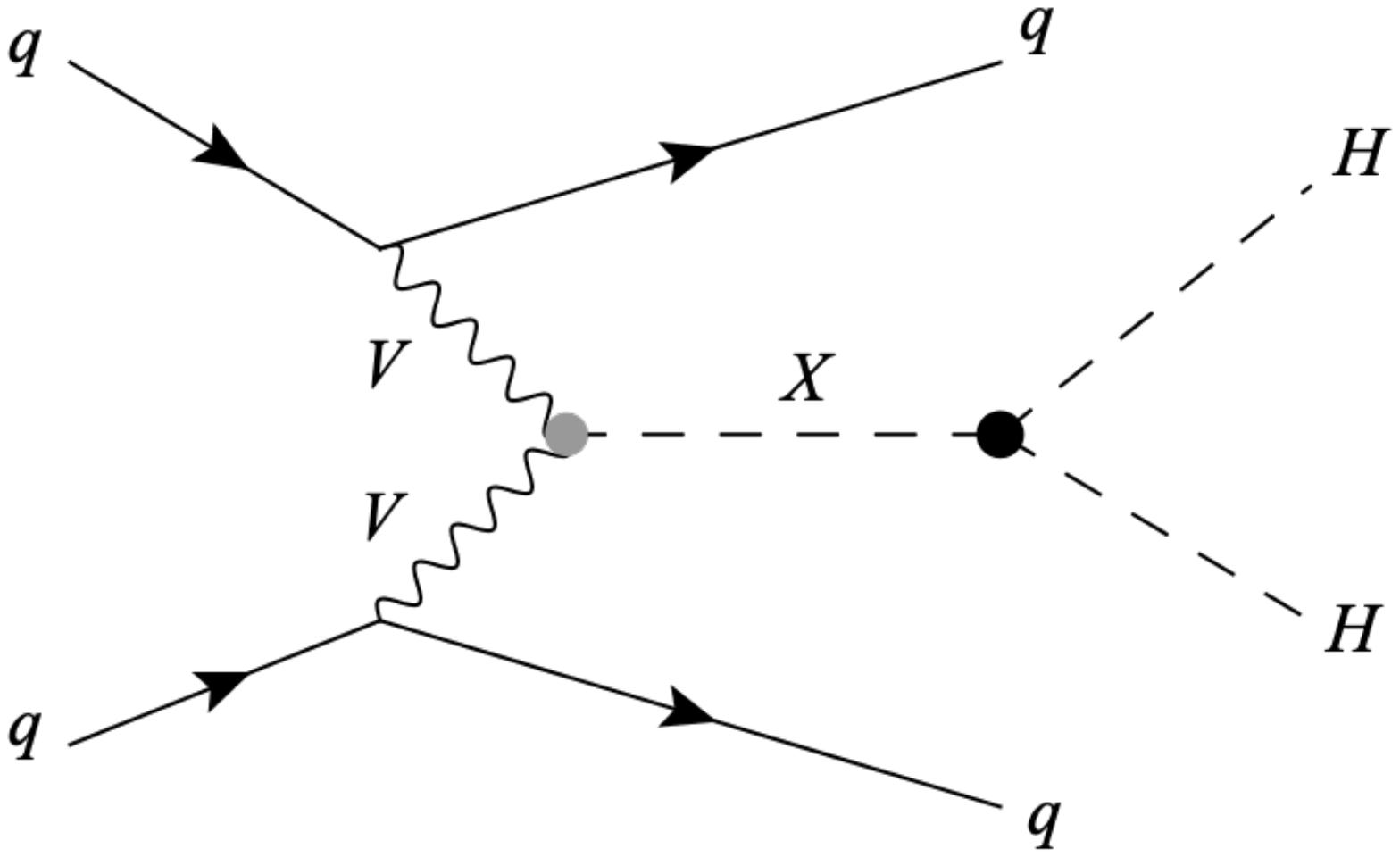}\label{fig:feynman_f}}
\caption{
Examples of leading-order Feynman diagrams for Higgs boson pair production.
For nonresonant ggF production, diagram \subref{fig:feynman_a} involves solely the top-quark Yukawa coupling, while diagram \subref{fig:feynman_b} involves the Higgs boson self-coupling.
For nonresonant VBF production, diagram \subref{fig:feynman_c} involves the self-coupling, diagram \subref{fig:feynman_d} involves solely the coupling to vector bosons, and diagram \subref{fig:feynman_e} involves the coupling between two Higgs bosons and two vector bosons.
Diagram \subref{fig:feynman_f} illustrates the resonant production mode.
}
\end{figure}

At the LHC, the coupling modifiers $\kvv = g_{HHVV}/g_{HHVV}^{\text{SM}}$ and $\kl = \lambda_{HHH}/\lambda_{HHH}^{\text{SM}}$ are studied via the production of Higgs boson pairs (\HH production).
In the SM, the main nonresonant \HH production modes are via the gluon--gluon fusion process (ggF), with a \xs of $\sigma^\text{SM}_\text{ggF} = 31.1^{+2.1}_{-7.2}$~fb at next-to-next-to-leading order in QCD and including an approximation of finite top-quark-mass effects~\cite{Dawson:1998py,Shao:2013bz,deFlorian:2013jea,deFlorian:2015moa,Borowka:2016ehy,deFlorian:2016spz,Grazzini:2018bsd,Bonciani:2018omm,Baglio:2018lrj,Heinrich:2019bkc,Baglio:2021}, and via the vector-boson fusion process (VBF), with a \xs of $\sigma^\text{SM}_\text{VBF} = 1.73\pm0.04$~fb at next-to-next-to-next-to-leading order in QCD~\cite{Baglio:2013,Ling:2014,Dreyer:2018}.
The VBF production \xs depends critically on the value of $\kvv$ due to destructive interference between \cref{fig:feynman_d,fig:feynman_e}.
For example, a value of $\kvv = 0$ leads to a \xs that is over 15 times the SM prediction.
The Higgs bosons produced in non-SM $\kvv$ scenarios are expected to be more energetic and more central in the detector on average relative to the SM $\kvv = 1$ scenario~\cite{Bishara_2017}, resulting in higher acceptances and selection efficiencies in this study.
The leading-order Feynman diagrams of the ggF and VBF \HH processes are shown in \cref{fig:feynman_a,fig:feynman_b,fig:feynman_c,fig:feynman_d,fig:feynman_e}.
In the SM, the processes depicted in \cref{fig:feynman_a,fig:feynman_b} interfere destructively.
Other coupling modifiers involved in these processes are of less interest in this analysis.
The ggF \HH production mode is sensitive to $\kl$ while the VBF \HH production mode is sensitive to both $\kl$ and $\kvv$.
Heavy resonances ($X$) beyond the SM may contribute to resonant \HH production~\cite{Branco:2011iw,Barman:2020ulr}, as exemplified via the VBF process in \cref{fig:feynman_f}.
The boosted VBF process provides a distinct signature for investigating these resonances, allowing exploration of uncharted phase space.
These \HH processes were studied with various decay final states by ATLAS and CMS, including
\bbbb\cite{EXOT-2016-31,HDBS-2018-18,HDBS-2018-41,HDBS-2019-29,CMS-HIG-20-005,CMS-B2G-22-003},
\bbyy\cite{HIGG-2016-15,HDBS-2018-34,CMS-HIG-19-018},
\bbtautau\cite{HIGG-2016-16-witherratum,HDBS-2018-40,HDBS-2019-27,CMS-HIG-20-010},
$b\bar{b} WW^*$~\cite{HIGG-2016-27},
$\gamma\gamma WW^*$~\cite{HIGG-2016-20},
$WW^* WW^*$~\cite{HIGG-2016-24},
$b\bar b \ell\ell\ell\ell$~\cite{CMS-HIG-20-004},
$b\bar b \ell\ell + \met$~\cite{HDBS-2019-02},
and their statistical combinations~\cite{HDBS-2018-58,HDBS-2022-03,CMS-HIG-22-001}.
No significant excess over the SM background has been observed to date.
The most stringent observed (expected) 95\% confidence level (CL) interval for the \kvv coupling modifier was found to be $0.62 < \kvv < 1.41$ ($0.66 < \kvv < 1.37$) by searching for nonresonant pair production of highly energetic Higgs bosons decaying into bottom quarks by the CMS Collaboration~\cite{CMS-B2G-22-003}.
In Ref.~\cite{HDBS-2018-18}, resonant VBF \HH production in the mass range of \SIrange{260}{1000}{\GeV} and non-resonant VBF \HH production in the non-boosted regime are searched for and no deviations from the background-only hypothesis are observed.
 
This paper reports a search for nonresonant and resonant VBF \hhtobbbb production using the full Run~2 ATLAS proton--proton (\pp) collision data sample with an integrated luminosity of $\SI{140}{\ifb}$.
The search focuses on a Lorentz-boosted topology, where two high-energy Higgs bosons each form a large-radius jet, referred to as a large-$R$ jet.
This topology is particularly sensitive to non-SM values of $\kvv$, and as such one of the goals of this analysis is to constrain $\kvv$.
Assuming the SM branching ratio of 58.2\% for \htobb~\cite{deFlorian:2016spz,Djouadi:1997yw}, approximately one third of \HH events lead to a \bbbb final state, making it the most abundant \HH final state.
A machine learning-based double \Pqb-tagging technique~\cite{ATL-PHYS-PUB-2020-019,ATL-PHYS-PUB-2021-035} uses the information from the large-$R$ jets and their constituents to identify \htobb decays.
The VBF signature is characterised by the presence of VBF jets that are defined as two small-$R$ jets with large invariant mass and rapidity separation.
This signature provides an effective handle for background suppression.
To maximise the sensitivity to the \kvv parameter, the nonresonant analysis is combined with the resolved analysis~\cite{HDBS-2019-29} where the four \Pqb-quarks are reconstructed as small-$R$ jets.
The Higgs bosons considered in the resolved analysis have lower transverse momentum (\pt) compared to those in this boosted search.
To avoid double counting events in the boosted nonresonant analysis presented in this paper, events that satisfy the resolved and boosted analysis selection are removed from the boosted analysis.
For the first time, a search for a new heavy spin-0 resonance that would mediate VBF Higgs boson pair production is carried out in the mass range of \SIrange{1}{5}{\TeV}.
 
The paper is structured as follows.
\Cref{sec:detector} briefly introduces the ATLAS detector.
\Cref{sec:mc} details the data and simulation samples used.
\Cref{sec:selection} describes the analysis selection.
\Cref{sec:bkg} explains the background estimate derived from data, and \cref{sec:bdt} covers the multivariate discriminants used.
Systematic uncertainties considered are detailed in \cref{sec:syst}.
Results are provided in \cref{sec:results}, and conclusions are given in \cref{sec:conclusion}.
 
\FloatBarrier
\section{The ATLAS experiment}
\label{sec:detector}
 
The ATLAS experiment~\cite{PERF-2007-01} at the LHC is a multipurpose particle detector with a forward--backward symmetric cylindrical geometry and a near \(4\pi\) coverage in solid angle.\footnote{
ATLAS uses a right-handed coordinate system with its origin at the nominal interaction point (IP) in the centre of the detector and the \(z\)-axis along the beam pipe.
The \(x\)-axis points from the IP to the centre of the LHC ring, and the \(y\)-axis points upwards.
Polar coordinates \((r,\phi)\) are used in the transverse plane, \(\phi\) being the azimuthal angle around the \(z\)-axis.
The pseudorapidity is defined in terms of the polar angle \(\theta\) as \(\eta = -\ln \tan(\theta/2)\) and is equal to the rapidity $ y = \frac{1}{2} \ln \left( \frac{E + p_z c}{E - p_z c} \right) $ in the relativistic limit.
Angular distance is measured in units of \(\Delta R \equiv \sqrt{(\Delta y)^{2} + (\Delta\phi)^{2}}\).
}
It consists of an inner tracking detector surrounded by a thin superconducting solenoid providing a \qty{2}{\tesla} axial magnetic field, electromagnetic and hadronic calorimeters, and a muon spectrometer.
The inner tracking detector covers the pseudorapidity range \(|\eta| < 2.5\).
It consists of silicon pixel, silicon microstrip, and transition radiation tracking detectors.
Lead/liquid-argon (LAr) sampling calorimeters provide electromagnetic (EM) energy measurements with high granularity within the region \(|\eta|< 3.2\).
A steel/scintillator-tile hadronic calorimeter covers the central pseudorapidity range (\(|\eta| < 1.7\)).
The endcap and forward regions are instrumented with LAr calorimeters for EM and hadronic energy measurements up to \(|\eta| = 4.9\).
The muon spectrometer surrounds the calorimeters and is based on three large superconducting air-core toroidal magnets with eight coils each.
The field integral of the toroids ranges between \num{2.0} and \qty{6.0}{\tesla\metre} across most of the detector.
The muon spectrometer includes a system of precision tracking chambers up to \(|\eta| = 2.7\) and fast detectors for triggering up to \(|\eta| = 2.4\).
The luminosity is measured mainly by the LUCID--2~\cite{LUCID2} detector, which is located close to the beampipe.
A two-level trigger system is used to select events~\cite{TRIG-2016-01}.
The first-level trigger is implemented in hardware and uses a subset of the detector information to accept events at a rate below \qty{100}{\kHz}.
This is followed by a software-based trigger that reduces the accepted event rate to \qty{1}{\kHz} on average depending on the data-taking conditions.
A software suite~\cite{ATL-SOFT-PUB-2021-001} is used in data simulation, in the reconstruction and analysis of real and simulated data, in detector operations, and in the trigger and data acquisition systems of the experiment.

\section{Data and simulation}
\label{sec:mc}
The analysis is performed using Run~2 ATLAS $pp$ collision data collected between 2015 and 2018.
The average number of interactions per proton bunch crossing (\pileup) is between 13 and 38 interactions, depending on the year~\cite{DAPR-2021-01}.
After applying ATLAS data quality requirements~\cite{DAPR-2018-01}, the data corresponds to an integrated luminosity of $\SI{140}{\ifb}$.
 
Monte Carlo (MC) simulation is used for the modelling of \HH processes, top-quark pair production (\ttbar) and multijet background processes.
The \ttbar and multijet samples are used solely for event selection optimisation and are identical to those used in Ref.~\cite{HDBS-2019-29}.
The final background estimate is obtained through data-driven techniques and described in \cref{sec:bkg}.
For all \HH samples, the Higgs boson mass is fixed to \SI{125}{\GeV}.
All samples are processed using the ATLAS simulation framework~\cite{SOFT-2010-01} where the detector response is simulated with \GEANT~\cite{Agostinelli:2002hh}.
The VBF \HH samples are simulated using \MADGRAPH[2.7.3]~\cite{Alwall:2014hca} at leading-order (LO) in quantum chromodynamics (QCD) with the \NNPDF[3.0nlo] parton distribution function (PDF) set~\cite{Ball:2014uwa}.
Samples with coupling modifier values $(\kl, \kvv, \kv)$ = $(1, 1, 1)$, $(1, 1.5, 1)$, $(2, 1, 1)$, $(10, 1, 1)$, $(1, 1, 0.5)$, $(-5, 1, 0.5)$, $(0, 1, 1)$, $(1, 0, 1)$, and $(1, 3, 1)$ are explicitly generated and a linear combination~\cite{HDBS-2019-29} of the first six of the listed samples is used to determine the expected yields and distributions for any value of $(\kl, \kvv, \kv)$.
The method is validated using the remaining simulated samples and good agreement is observed.
The SM ggF \HH samples are simulated using the \POWHEGBOX[v2] generator~\cite{Nason:2004rx, Frixione:2007vw, Alioli:2010xd} at next-to-leading-order (NLO) in QCD, including finite top-quark-mass effects, using the \PDFforLHC[15]~\cite{Butterworth:2015oua} PDF set.
A reweighting technique based on the particle-level invariant mass \mhh of the Higgs boson pair is applied to the $\kl = 1$ sample to determine the ggF \HH yield and kinematic distributions for any value of $\kl$~\cite{ATL-PHYS-PUB-2019-007}.
The ggF \HH samples are considered as background processes when constraining \kvv and as signal processes when deriving the results related to \kl.
 
The hypothetical heavy spin-0 resonance $X$ that would mediate VBF \HH, $pp \to X+jj \to HH+jj$, is simulated using the \MGNLO[2.6.1]~\cite{Alwall:2014hca} generator at LO in QCD with the \NNPDF[2.3lo]~\cite{Ball:2012cx} PDF set.
The branching ratio of $X \to \HH$ is set to 100\%.
Two resonance-width hypotheses are considered, where the resonance width is denoted by $\Gamma_{X}$: a generic narrow-width signal ($\Gamma_{X}$ smaller than the detector resolution of \SIrange{5}{6}{\%} of the resonance mass) and a broad-width signal ($\Gamma_{X} = \SI{20}{\%}$ of the resonance mass) based on the Composite Higgs model~\cite{Contino_2011}.
These samples cover a range of resonance masses, denoted by \mx, from \SI{1}{\TeV} to \SI{5}{\TeV}, with increased spacing between the higher mass points and a different number of points between the narrow- and broad-width assumptions.
For all resonant and nonresonant \HH samples, parton showers and hadronisation are simulated using \PYTHIA[8.244]~\cite{Sjostrand:2014zea} with the A14 set of tuned parameters~\cite{ATL-PHYS-PUB-2014-021} and the \NNPDF[2.3lo] PDF set.
\EVTGEN[1.7.0]~\cite{Lange:2001uf} is used to model the properties of heavy-flavour decays.
The effects of \pileup are modelled by superimposing each simulated hard-scattering event with inelastic \pp events simulated using \PYTHIA[8.186]~\cite{Sjostrand:2007gs} with the \NNPDF[2.3lo] PDF set and the A3 set of tuned parameters~\cite{ATL-PHYS-PUB-2016-017}, and is weighted to match the observed \pileup in data.
 
\section{Event selection}
\label{sec:selection}
Events must satisfy trigger decisions that require minimum transverse energies of the triggered large-$R$ jet.
The threshold varies between \SIrange{360}{420}{\GeV}, depending on the year of data taking~\cite{TRIG-2016-01,ATL-DAQ-PUB-2019-001,TRIG-2019-02}.
Events are required to contain two Higgs boson candidates and two VBF jets.
A Higgs boson candidate is reconstructed as a large-$R$ jet, denoted by $J$, using the anti-\kt algorithm~\cite{Cacciari:2008gp, Fastjet} with the radius parameter $R=1.0$.
The large-$R$ jets are reconstructed from topological energy depositions~\cite{PERF-2014-07} in the calorimeter and are trimmed~\cite{Krohn:2009th,JETM-2018-06} to improve the jet mass resolution and to mitigate the effects of \pileup and soft radiation.
A method similar to the one used in Ref.~\cite{HDBS-2019-29} is used to correct the four-momentum of large-$R$ jets by accounting for energy lost to soft out-of-cone radiation and to muons and neutrinos in semileptonic $b$-hadron decays.
This correction improves the jet mass resolution.
The mass of a large-$R$ jet ($m_J$) is calculated using a combination of calorimeter and tracking information~\cite{ATLAS-CONF-2016-035} to improve the resolution over the whole range of jet \pT.
The large-$R$ jets must satisfy $\SI{250}{\GeV} < \pT < \SI{3000}{\GeV}$, $|\eta| < 2.0$ and $\SI{50}{\GeV} < m_J < \SI{600}{\GeV}$, corresponding to the region where the jet calibration is valid.
The two leading \pT large-$R$ jets are considered as the Higgs boson candidates, and the leading jet \pT criterion is raised to $\pT > \SI{450}{\GeV}$ to ensure that the online trigger is fully efficient.
The leading (\hig{1}) and sub-leading (\hig{2}) Higgs boson candidates are ordered by their \pt.
A double \Pqb-tagging algorithm based on a deep neural network~\cite{ATL-PHYS-PUB-2020-019,ATL-PHYS-PUB-2021-035} is applied to the large-$R$ jets to identify \htobb decays.
The algorithm is trained on large-$R$ jets with masses above \SI{50}{\GeV} and calibrated using a $Z\to b\bar{b}$ control sample in four $\pt^{J}$ regions.
When neglecting systematic uncertainties, compared to using variable-radius track-jet $b$-tagging~\cite{ATL-PHYS-PUB-2017-010,ATL-PHYS-PUB-2017-013}, the algorithm provides a sensitivity improvement of up to \SI{50}{\%} in expected discovery significance for the $H\to b \bar{b}$ analysis~\cite{EXOT-2018-46}.
Events with two Higgs boson candidates satisfying the \SI{60}{\%} efficiency working point are referred to as \textsc{2Pass} events.
This working point reduces multijet (top-quark) events by a factor of 92 (31).
Events with only one Higgs boson candidate satisfying the \SI{60}{\%} efficiency working point are referred to as \textsc{1Pass} and are used for background estimation.
The small-$R$ jets, denoted by $j$, are reconstructed from particle-flow objects~\cite{PERF-2015-09} using the anti-\kt algorithm with $R=0.4$.
The jet energy ($E$) is corrected by applying \textit{in situ} corrections that reduce the contribution from \pileup jets~\cite{JETM-2018-05}.
The small-$R$ jets must have $\pT > \SI{20}{\GeV}$ and $|\eta| < 4.5$, and those with $\pT < \SI{60}{\GeV}$ and $|\eta| < 2.4$ must satisfy a requirement based on the output of the multivariate jet vertex tagger algorithm~\cite{PERF-2014-03} to reduce the effect from \pileup.
To remove overlap with the Higgs boson candidates, the distance between a small-$R$ jet and the selected Higgs boson candidates must satisfy $\Delta R (J, j) > 1.4$.
The two leading \pT small-$R$ jets are assigned as VBF jets and required to satisfy the criteria $|\Delta\eta(j, j)| > 3$ and $m_{jj} > \SI{1}{\TeV}$.
 
\begin{figure}[htb]
\centering
\subfloat[]{\includegraphics[width=0.5\textwidth]{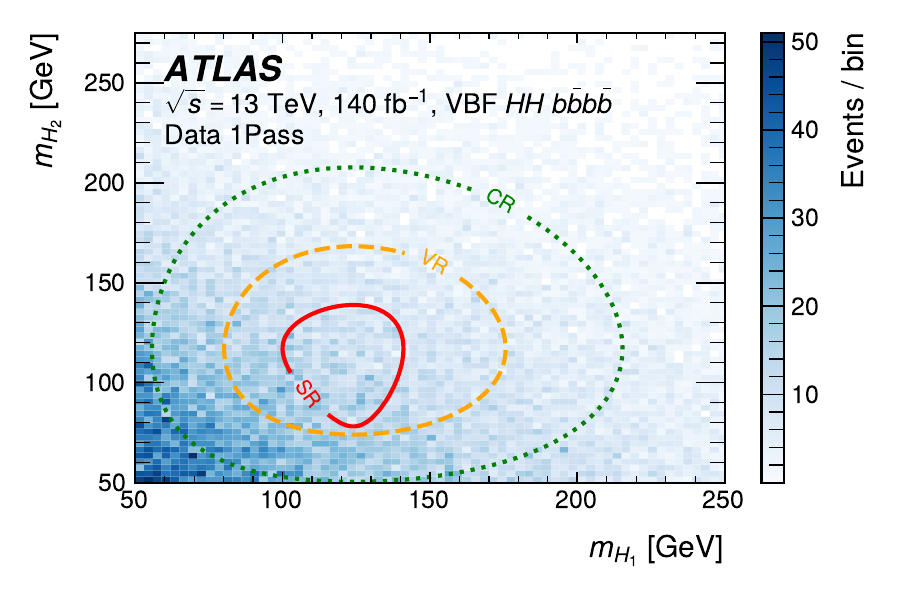}}
\subfloat[]{\includegraphics[width=0.5\textwidth]{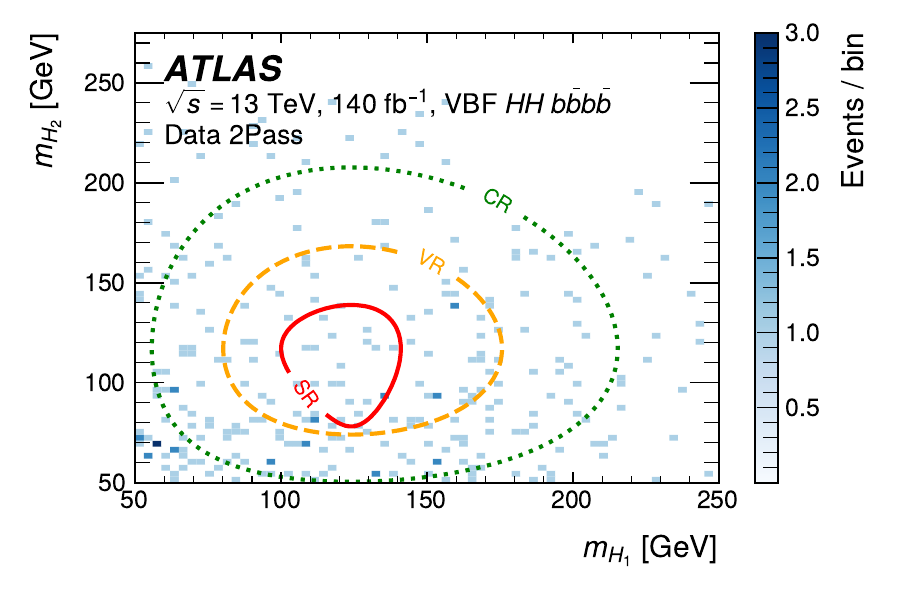}} \\
\subfloat[]{\includegraphics[width=0.5\textwidth]{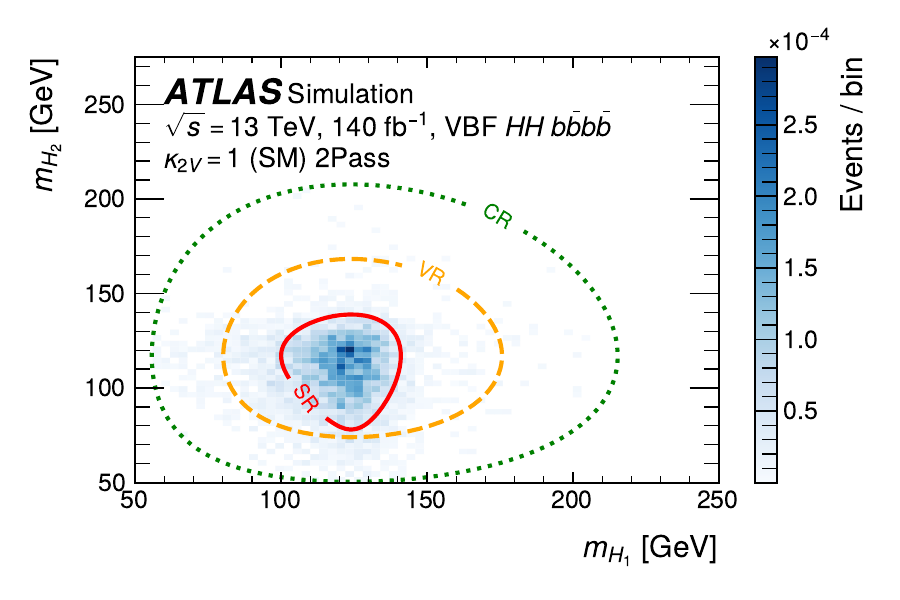}}
\subfloat[]{\includegraphics[width=0.5\textwidth]{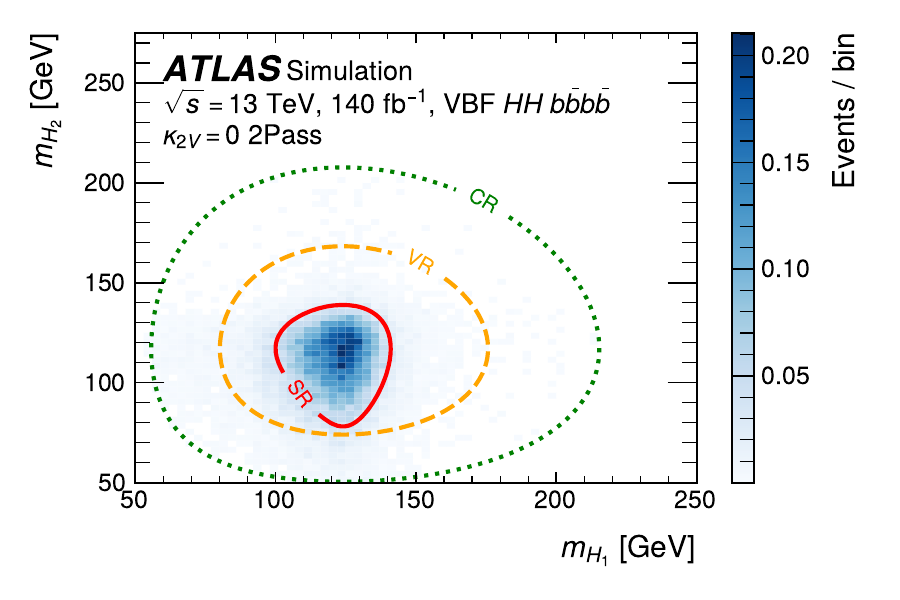}} \\
\subfloat[]{\includegraphics[width=0.5\textwidth]{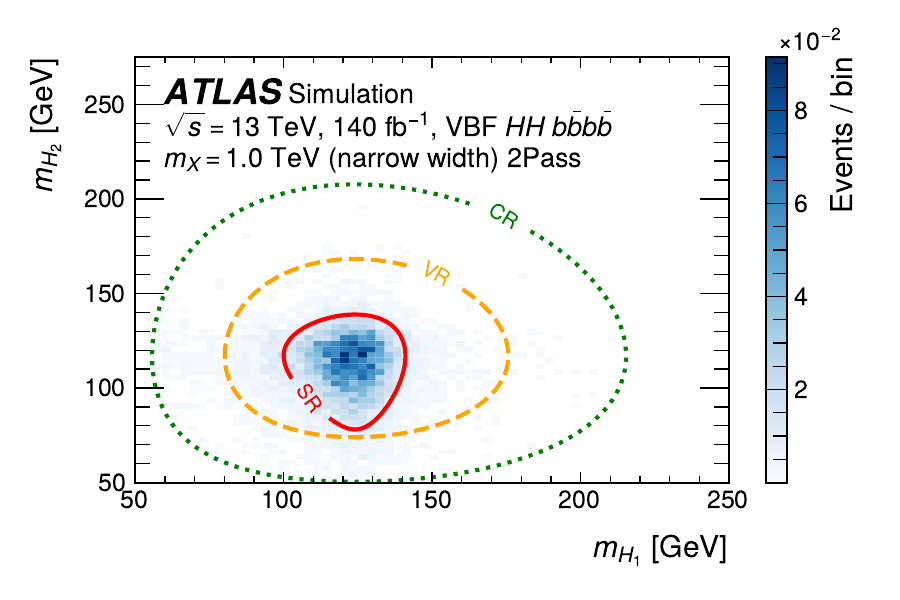}}
\caption{
The mass planes of the reconstructed Higgs boson candidates for the (a) \textsc{1Pass} and (b) \textsc{2Pass} selections of the analysis, shown for the data events.
The mass planes for the \textsc{2Pass} selection of the analysis are shown for the (c) VBF SM $\kvv = 1$ \HH, (d) VBF $\kvv = 0$ \HH, and (e) $\mx = \SI{1}{\TeV}$ spin-0 narrow-width resonance \HH samples.
The continuous red line describes the Signal Region (SR).
The Validation Region (VR) lies between the dashed yellow line and the continuous red line.
The Control Region (CR) lies between the dotted green line and the dashed yellow line.
The bin sizes are \SI{1.33}{\GeV} by \SI{1.33}{\GeV}.
}
\label{fig:massplane}
\end{figure}

After the preselections described above, \textsc{1Pass} and \textsc{2Pass} events are separately classified into signal regions (SRs), validation regions (VRs), and control regions (CRs) according to the following criteria defined in the \mhhplane.
The SR, VR, and CR are disjoint by construction such that there is no overlap between them.
Events in the SR reside in the region defined by
\begin{equation}
\sqrt{\left(\frac{\mh{1} - \SI{124}{\GeV}}{\SI{1500}{\GeV} / \mh{1}}\right)^{2} + \left(\frac{\mh{2} - \SI{117}{\GeV}}{\SI{1900}{\GeV} / \mh{2}}\right)^{2}} < \SI{1.6}{\GeV}.
\label{eqn:SR}
\end{equation}
Events in the VR reside in the region bounded by the SR boundary and
\begin{equation}
\sqrt{\left(\frac{\mh{1} - \SI{124}{\GeV}}{0.1 \ln(\mh{1})}\right)^{2} + \left(\frac{\mh{2} - \SI{117}{\GeV}}{0.1 \ln(\mh{2})}\right)^{2}} < \SI{100}{\GeV},
\label{eqn:VR}
\end{equation}
and events in the CR reside in the region bounded by the VR outer boundary and
\begin{equation}
\sqrt{\left(\frac{\mh{1} - \SI{124}{\GeV}}{0.1 \ln(\mh{1})}\right)^{2} + \left(\frac{\mh{2} - \SI{117}{\GeV}}{0.1 \ln(\mh{2})}\right)^{2}} < \SI{170}{\GeV}.
\label{eqn:CR}
\end{equation}
The variables \mh{1} and \mh{2} in these equations are in units of \GeV.
The values of \SI{124}{\GeV} and \SI{117}{\GeV} in Eqs.~(\ref{eqn:SR})--(\ref{eqn:CR}) are chosen such that they correspond to the centres of the \mh{1} and \mh{2} distributions of the VBF \HH events from simulation.
These centres deviate from the measured Higgs boson mass of~\SI{125}{\GeV} due to detector effects, and energy lost to neutrinos from the $b$-hadron decays and to out-of-cone radiation.
The SR definition is optimised to maximise the overall $S/\sqrt{B}$.
The signal $S$ is the yield of $\kvv=0$ VBF \HH events in simulation which is chosen to maximise the sensitivity to the $\kvv$ coupling as it is a representative proxy for non-SM $\kvv$ samples.
The background $B$ is the expected number of background events estimated by using the \ttbar and multijet simulated samples.
As multijet background processes preferentially populate the \mhhplane in the lower Higgs boson candidate mass region compared to \HH processes, Eqs.~(\ref{eqn:VR}) and~(\ref{eqn:CR}) help reduce contributions from multijet events.
The boundaries of the SR, VR, and CR in the reconstructed \mhhplane are shown in \cref{fig:massplane} for the \textsc{1Pass} and \textsc{2Pass} selections of the analysis.
The \mhhplane is smoothly falling across the Higgs boson candidate masses.
Most \HH events are captured by the signal region boundary; the fraction of \textsc{2Pass} events in the SR is 76\% (78\%--55\%) for nonresonant (resonant 1~\TeV--5~\TeV) events.
The overall signal acceptance times efficiency in the \textsc{2Pass} SR ranges from 1\% for a representative nonresonant non-SM signal sample to 0.02\% for the SM nonresonant signal sample due to different kinematics.
For the resonant signal samples, the overall acceptance times efficiency ranges from 5\% to 10\%, depending on the mass and width of the resonance.
 
\FloatBarrier
\section{Background modelling}
\label{sec:bkg}
Background processes in the SR predominantly originate from nonresonant multijet production of multiple heavy ($\Pqb/\Pqt$) quarks and from jets initiated by non-heavy quarks misidentified as originating from heavy quarks.
The background contribution coming from single-Higgs boson and diboson events were found to be negligible.
The multijet background, which is composed of approximately 10\% \ttbar events, is estimated by using a data-driven method and \textsc{1Pass} events.
The signal contamination in the \textsc{1Pass} selection is verified to be at most \SI{1}{\%} overall, for any signal considered.
In the most signal-like bin of the final discriminant (described in \cref{sec:bdt}), it increases to up to 8\%, which is below the statistical uncertainty of this bin.
As the difference between the shape of the final discriminant in \textsc{1Pass} and \textsc{2Pass} events is within statistical uncertainty,
\ifdoCONF
\else
as shown in \cref{fig:VR1P2P},
\fi
an inclusive normalisation factor is derived from the CR and applied to the SR\@.
The normalisation factor $w$ is derived by calculating the ratio of the number of events in the CR \textsc{2Pass} and CR \textsc{1Pass}: $w = 0.0081 \pm 0.0010$.
 
The uncertainty is obtained by re-deriving this ratio in the VR and computing the difference between the value of $w$ derived in the CR and the VR, and is used as an overall uncertainty in the multijet yield in the fit described in \cref{sec:results}.
The background estimate in the SR \textsc{2Pass} is thus obtained by multiplying the relevant distribution in the SR \textsc{1Pass} by $w$.
Alternative definitions for the CR and VR boundaries, which split the nominal definitions of CR and VR into quadrants, are found to yield values of $w$ that are consistent with the nominal estimate.
To cover any potential residual shape differences, a shape uncertainty in the final discriminant is estimated by taking the relative difference between the \textsc{1Pass} and \textsc{2Pass} discriminant distributions in the VR and symmetrising around the background estimate.
\ifdoCONF
\else
\begin{figure}[htbp]
\centering
\includegraphics[width=0.55\linewidth]{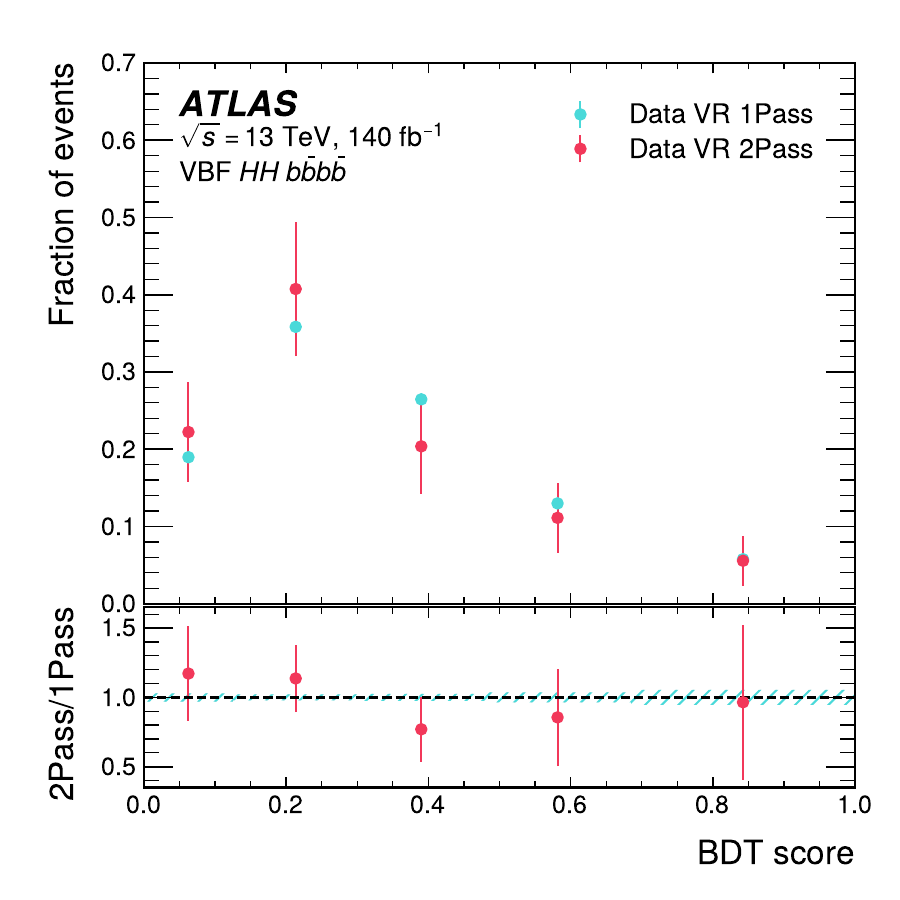}
\caption{
The BDT score (see \cref{sec:bdt}) for data events in the VR for the \textsc{2Pass} and \textsc{1Pass} selections.
The blue histogram is the distribution in VR \textsc{1Pass}.
The red histogram is the distribution in VR \textsc{2Pass}.
The lower panel shows the ratio of the \textsc{2Pass} and \textsc{1Pass} selections.
The error bars on the data points represent the statistical uncertainty.
}
\label{fig:VR1P2P}
\end{figure}
\fi
 
\FloatBarrier
\section{Multivariate discriminants}
\label{sec:bdt}
Boosted decision trees (BDTs) implemented in the XGBoost~\cite{Chen:2016:XST:2939672.2939785} library are used to separate signal events from background events in the SR\@.
In both the nonresonant and resonant analyses, orthogonality between training, testing, and validation samples is ensured by splitting the available data by event number.
Hyperparameters are optimised using the validation samples to enhance the classifier's performance, and the kinematic variables used as input are listed in \cref{tab:BDT_inputs_short}.
In the nonresonant analysis, a BDT is trained to separate $\kvv = 0$ signal events from background events consisting of the nonresonant multijet background estimate and SM ggF and VBF \HH production events.
The $\kvv = 0$ signal is chosen as a representative proxy for non-SM values of $\kvv$, allowing to maximise the sensitivity to the $\kvv$ coupling.
 
The resonant analysis uses a mass-parameterised BDT (pBDT) to accommodate multiple resonant signals with different mass hypotheses, inspired by parameterised neural networks~\cite{Baldi_2016}.
In addition to the variables listed in \cref{tab:BDT_inputs_short}, the pBDT includes the truth mass of the heavy resonances as an additional input parameter.
Signals are composed of thirteen narrow-width MC samples with distinct hypotheses for the truth mass of the heavy resonance.
The broad-width samples are not used during training.
The background is taken from the data-driven estimate described in \cref{sec:bkg}.
A random value from the available signal true resonance masses \mx is assigned to each background event in the training.
To ensure an adequate number of training events, the requirements on the VBF jets are removed and the double \Pqb-tagging working point requirements are relaxed to the \SI{70}{\%} efficiency working point during training for the resonant analysis.
The nominal selection is reinstated after training.
 
\begin{table}[tbh]
\centering
\renewcommand{\arraystretch}{1.3}
\caption{
Kinematic variables used in the BDT training in both the nonresonant and resonant analyses.
Additionally, the truth mass of the resonance is used as an input variable in the resonant analysis.
}
\begin{tabular}{l|ll}
\toprule
Physics objects & BDT input variables \\
\hline
Higgs boson candidate ($H_i$, $i=1,2$) & $\pT^{H_i}$, $\eta_{H_i}$             \\
Di-Higgs boson system ($HH$)                 & $\pT^{HH}$, $\eta_{HH}$, $m_{HH}$     \\
VBF jets ($j_i$, $i=1,2$)              & $\pT^{j_i}$, $\eta_{j_i}$, $E_{j_i}$  \\
\bottomrule
\end{tabular}
\label{tab:BDT_inputs_short}
\end{table}
 
\FloatBarrier
\vspace{-0.3cm}
\section{Systematic uncertainties}
\label{sec:syst}
Systematic uncertainties in the background and signals are evaluated for a variety of sources.
Both a normalisation and shape uncertainty are assigned to the data-driven background estimate, as described in \cref{sec:bkg}.
Uncertainties resulting from detector effects only affect signal simulation.
The impact of the main sources of uncertainty in the signal yield are evaluated for various hypothesised signals.
The dominant systematic uncertainty stems from the double \Pqb-tagging algorithm (\SIrange{20}{30}{\%}).
It is derived in four $\pt^{J}$ bins using a $Z\to b\bar{b}$ control sample~\cite{ATL-PHYS-PUB-2021-035}.
As this $Z\to b\bar{b}$ control sample is statistically limited, the systematic uncertainty coming from the double \Pqb-tagging algorithm is uncorrelated across the four $\pt^{J}$ bins.
The uncertainty in the integrated luminosity is \SI{0.83}{\%}~\cite{DAPR-2021-01}.
The uncertainty in the \pileup modelling is $<0.1\%$.
Uncertainties affecting the final state reconstruction and identification include the energy and mass scales of the large-$R$ jets (\SIrange{1}{10}{\%}), the large-$R$ jet energy resolution and mass resolution ($<1\%$)~\cite{JETM-2018-02,ATLAS-CONF-2020-022}, and the small-$R$ jet energy scale and resolution (\SIrange{1}{10}{\%})~\cite{PERF-2016-04,JETM-2018-05}.
The efficiency and acceptance of nonresonant and resonant signals are also affected by theoretical modelling uncertainties, such as the parton showering (\SIrange[range-phrase = --]{5}{10}{\%}) and renormalisation and factorisation scale choices (\SIrange[range-phrase = --]{1}{5}{\%}).
Theoretical uncertainties in the \htobb branching ratio (\SI{3.5}{\%})~\cite{deFlorian:2016spz} are included.
Theoretical uncertainties in the nonresonant ggF and VBF \HH \xss arising from uncertainties in the PDF and \alphas, and the choice of renormalisation scheme and the scale of the top-quark mass, are taken from Refs.~\cite{deFlorian:2016spz,Baglio:2021,Grazzini:2018bsd}.
No theoretical uncertainties in the resonant \HH \xss are considered.
The analysis is ultimately limited by statistical uncertainties.
 
\FloatBarrier
\vspace{-0.2cm}
\section{Results}
\label{sec:results}
\begin{figure}[htbp]
\centering
\includegraphics[width=\linewidth]{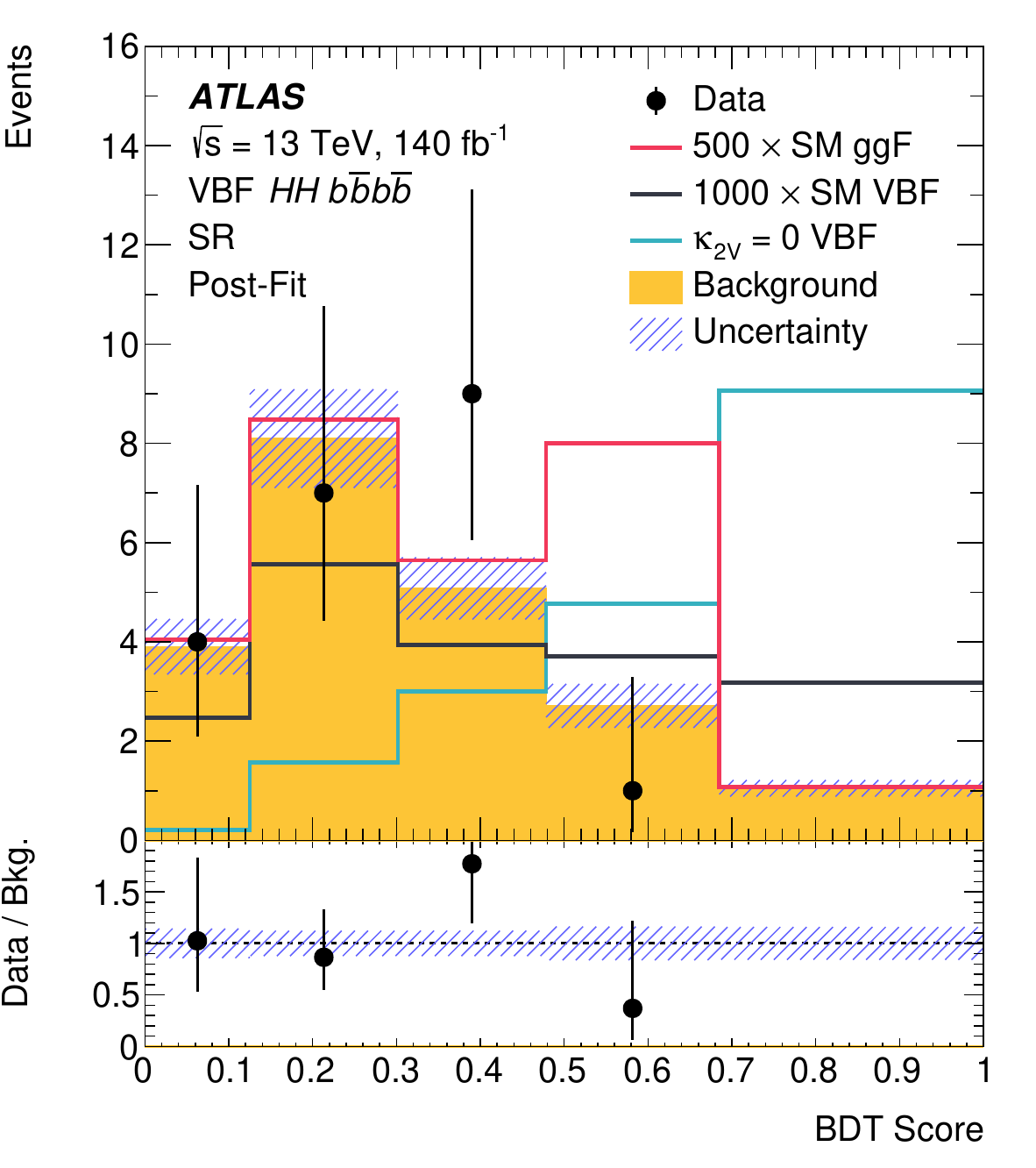}
\caption{
The distribution of the BDT score used in the nonresonant analysis after a background-only fit to the data in the signal region.
The distributions corresponding to the SM ggF \HH, SM VBF \HH, and $\kvv=0$ VBF \HH samples are also shown, in some cases scaled by a factor for visibility.
The lower panel shows the ratio of data to the total background prediction, with its uncertainty represented by the shaded band.
The error bars on the data points represent the statistical uncertainty.
}
\label{fig:postfit}
\end{figure}

A binned maximum-likelihood fit to the BDT distributions in the \textsc{2Pass} SR is carried out with the systematic uncertainties parameterised as nuisance parameters.
The BDT output binning transformation is the same as the one detailed in Ref.~\cite{HIGG-2018-51}.
The observed BDT distribution of data, and the background-only fit to the distribution, is presented in \cref{fig:postfit}.
Good agreement is found between the data and the background-only hypothesis.
No data are observed in the most signal-like bin while the expectation from background contributions before the fit is $1.1\pm 0.2$ events ($1.0\pm 0.2$ events after the fit).
 
\begin{figure}[htb]
\centering
\subfloat[]{\includegraphics[width=.8\textwidth]{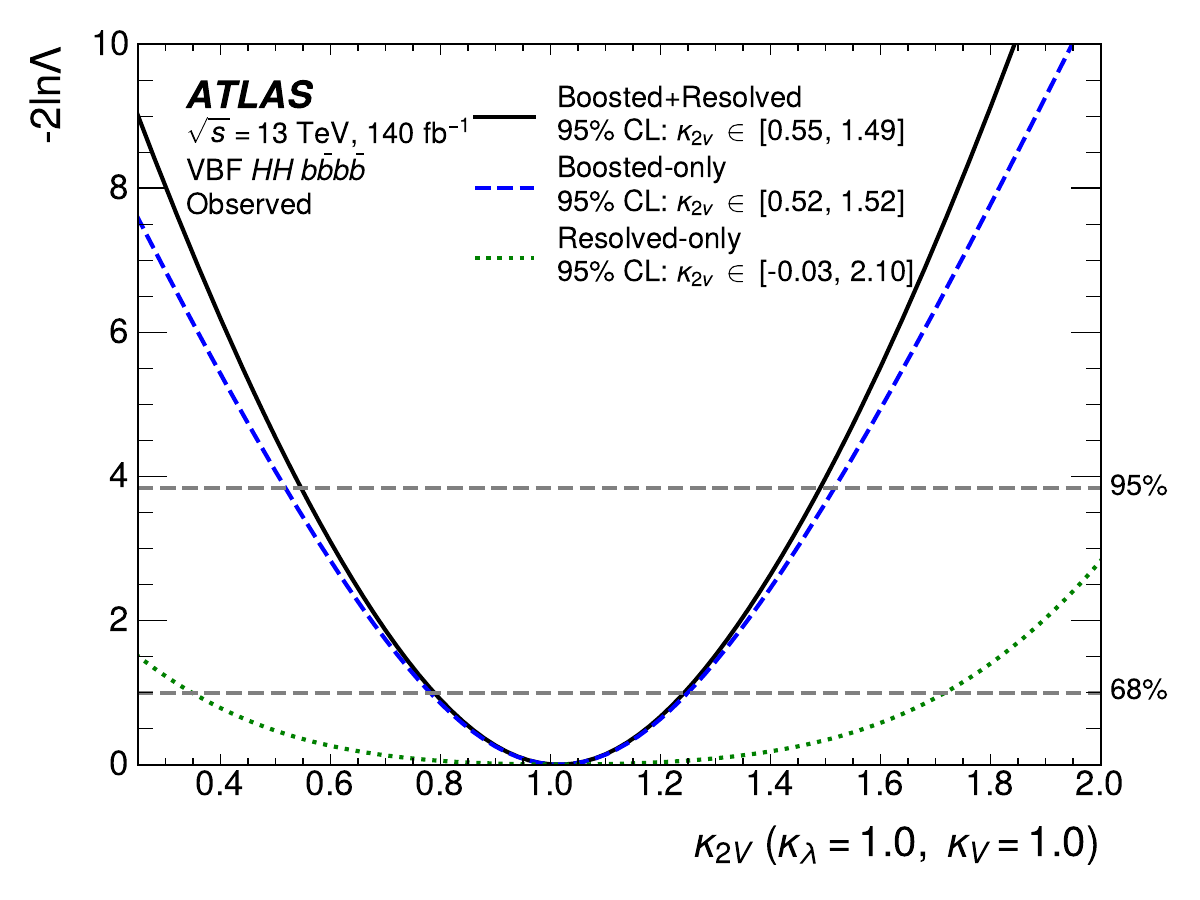}}\\
\subfloat[]{\includegraphics[width=.8\textwidth]{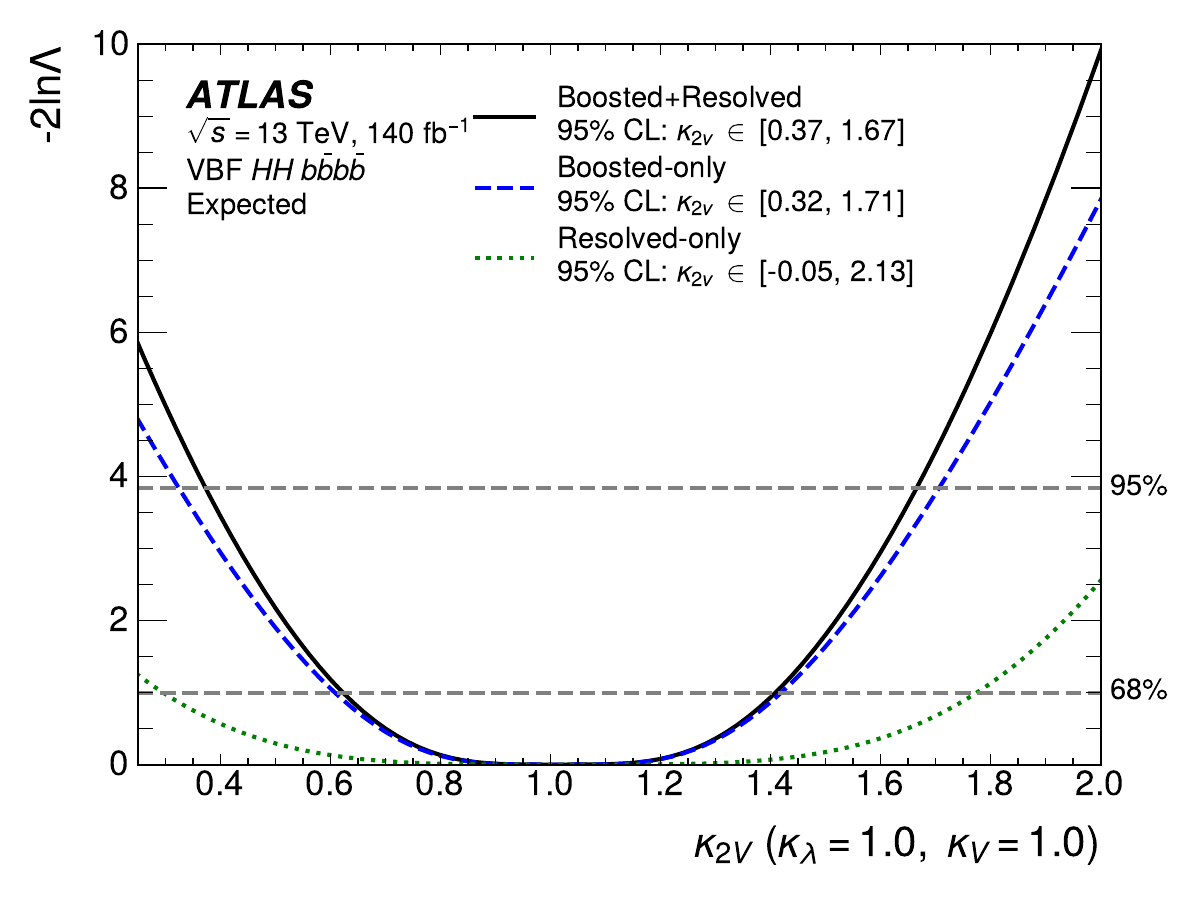}}
\caption{
Observed (a) and expected (b) values of $-2\ln\Lambda$ as a function of \kvv for the resolved (dotted green) and boosted (dashed blue) analyses, and their combination (solid black), with all other coupling modifiers fixed to their SM predictions.
}
\label{fig:likelihood_scan}
\end{figure}

\begin{figure}[htb]
\centering
\subfloat[]{\includegraphics[width=.79\textwidth]{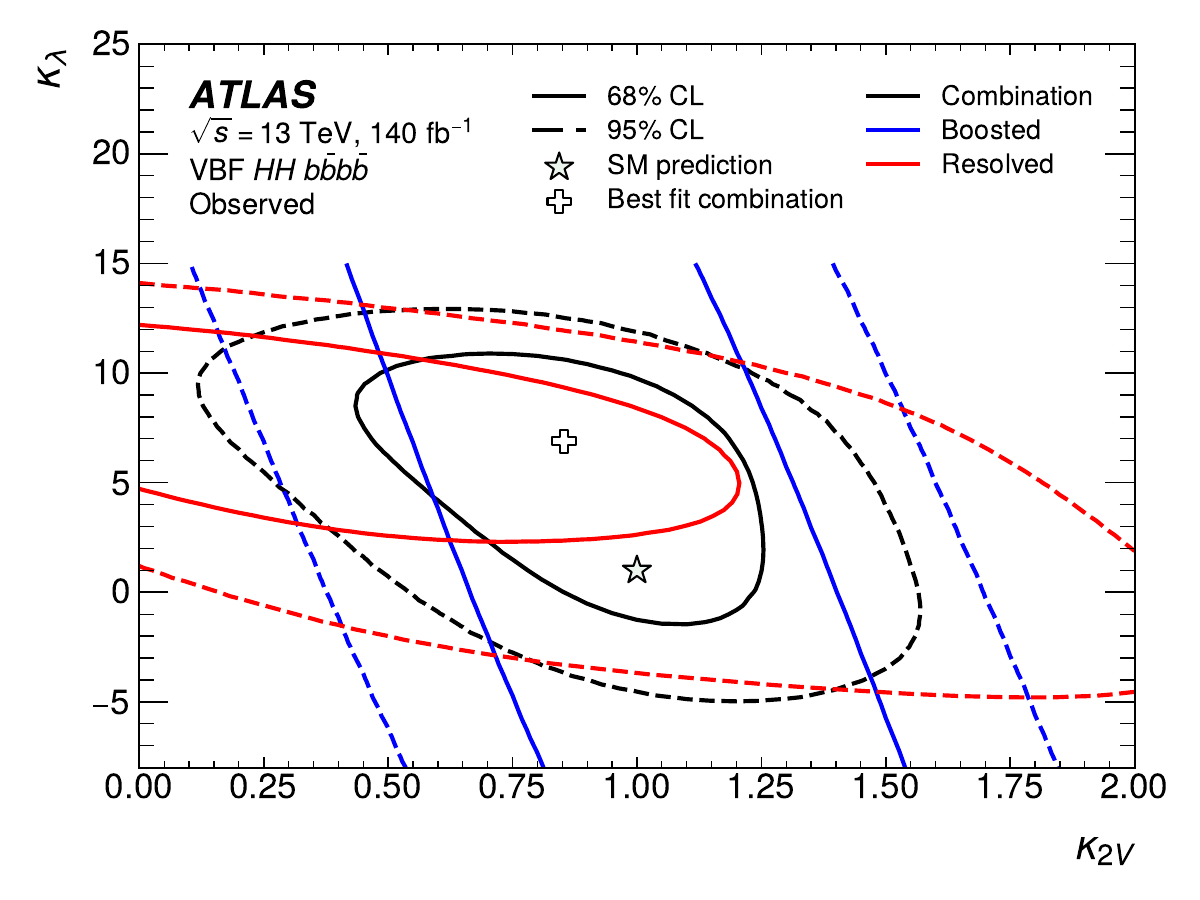}}\\
\subfloat[]{\includegraphics[width=.79\textwidth]{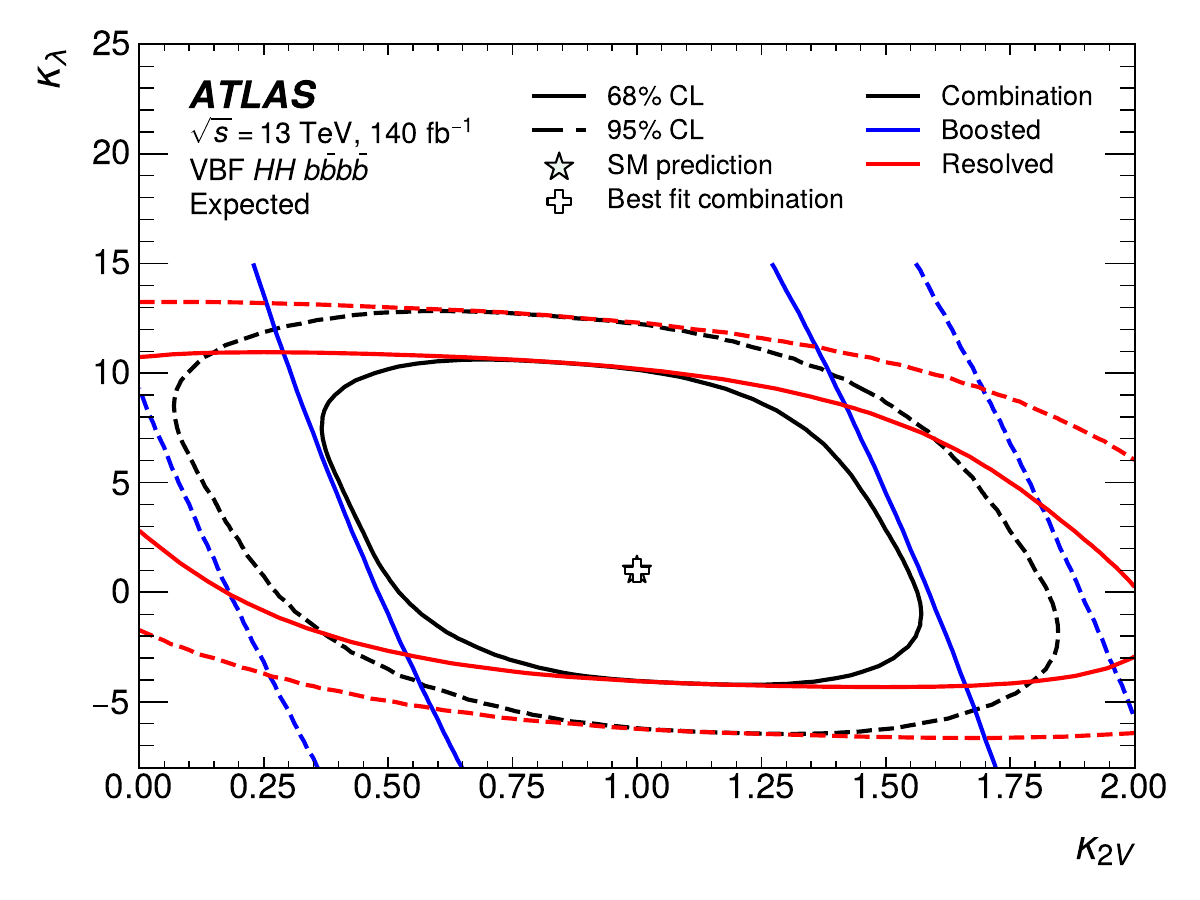}}
\caption{
Observed (a) and expected (b) likelihood contours at the 68\% (solid line) and 95\% (dashed line) CL in the $\kl$--$\kvv$ plane.
The red, blue, and black colours represent the resolved-only, boosted-only, and boosted+resolved combination results, respectively.
All other coupling modifiers are fixed to their SM predictions.
The SM prediction is indicated by the star, while the best-fit value is denoted by the cross.
The shift in the observed value from the SM prediction is driven by the resolved analysis.
The observed constraint on \kl values from the combination is less stringent than the constraint from the resolved-only fit due to the different best-fit values of the \kvv modifier.
The result for \kl values above 15 is not plotted for clarity.
}
\label{fig:likelihood_scan_2d}
\end{figure}

In the nonresonant search, a combination with the ggF and VBF categories of the resolved analysis~\cite{HDBS-2019-29} is additionally performed to improve the sensitivity to \kvv.
Uncertainties stemming from common sources in both the analyses are correlated.
The values of twice the negative-logarithm of the profile likelihood ratio ($-2\ln\Lambda$) as a function of \kvv are shown in \cref{fig:likelihood_scan} for the resolved and boosted analyses, and their combination.
The best-fit \kvv value obtained from the fit to the data is ${\kvvbestfitboostedonly}^{+\kvvbestfiterrhighboostedonly}_{-\kvvbestfiterrlowboostedonly}$ for the boosted result and ${\kvvbestfit}^{+\kvvbestfiterrhigh}_{-\kvvbestfiterrlow}$ for the combined result.
The boosted result provides an observed (expected) constraint of $\kvvobslowboostedonly < \kvv < \kvvobshighboostedonly$ ($\kvvexplowboostedonly < \kvv < \kvvexphighboostedonly$) at the \SI{95}{\%} CL\@.
The observed constraints on \kvv are stronger than expected due to the deficit of data events in the most signal-like bin.
The combined observed (expected) constraints obtained are $\kvvobslow < \kvv < \kvvobshigh$ ($\kvvexplow < \kvv < \kvvexphigh$) at the \SI{95}{\%} CL\@.
The allowed range of \kvv values is reduced by a factor of two compared with previous ATLAS publications~\cite{HDBS-2019-29}.
The Higgs boson coupling $\kvv = 0$ is excluded with an observed (expected) significance of $\obssignificanceboostedonly$ ($\expsignificanceboostedonly$) standard deviations.
When combining the boosted and resolved results, the Higgs boson coupling $\kvv = 0$ is excluded with an observed (expected) significance of $\obssignificance$ ($\expsignificance$) standard deviations.
These results are obtained assuming \kl and all other coupling values are as predicted by the SM\@.
The expected improvement in constraining \kl when including the boosted result is found to be marginal compared to the resolved-only result.
The exclusion constraints in the two-dimensional $\kl$--$\kvv$ coupling modifier space are presented in \cref{fig:likelihood_scan_2d}.
The resolved and boosted analyses are sensitive to complementary coupling parameters; the \kl sensitivity is driven by the resolved analysis, while the \kvv sensitivity is dominated by the boosted analysis.
 
\begin{figure}[htb]
\centering
\subfloat[]{\includegraphics[width=0.8\linewidth]{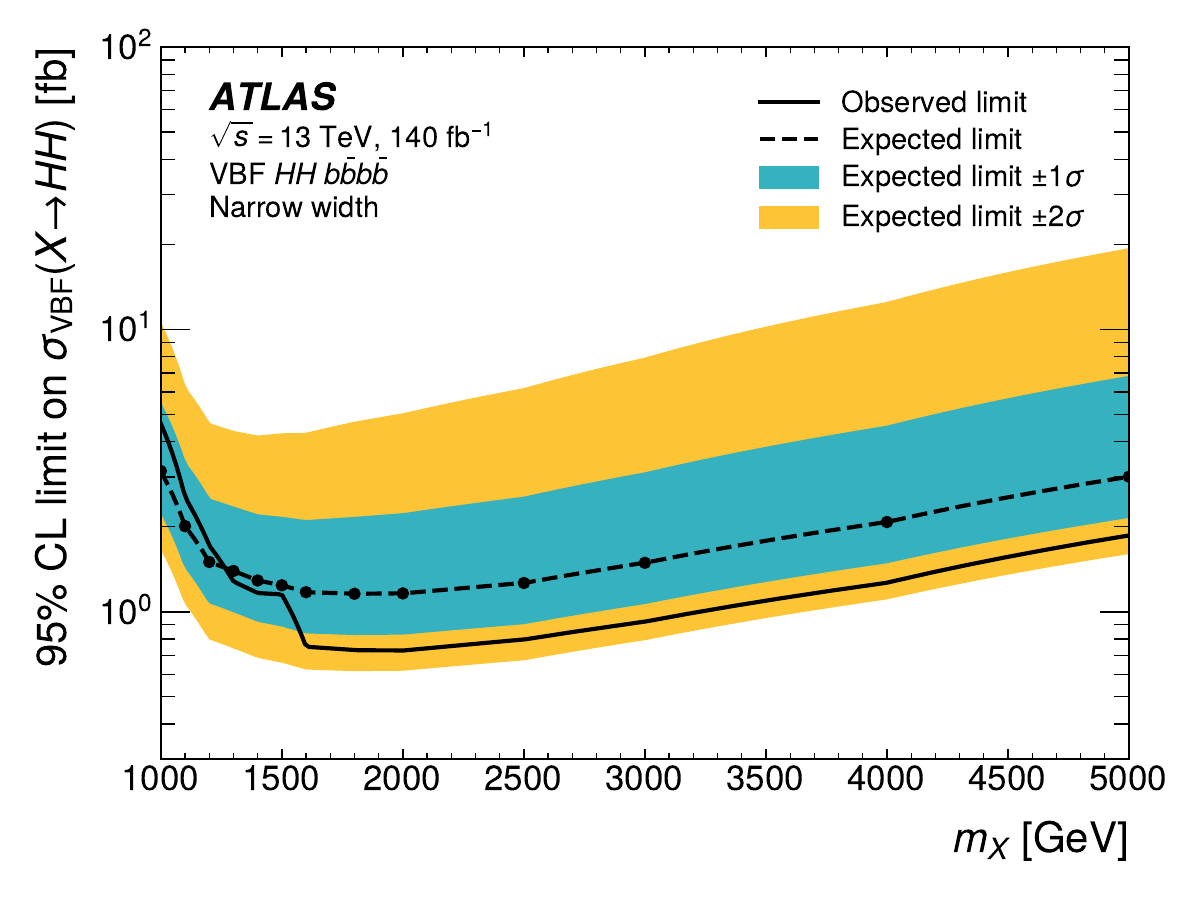}} \\
\subfloat[]{\includegraphics[width=0.8\linewidth]{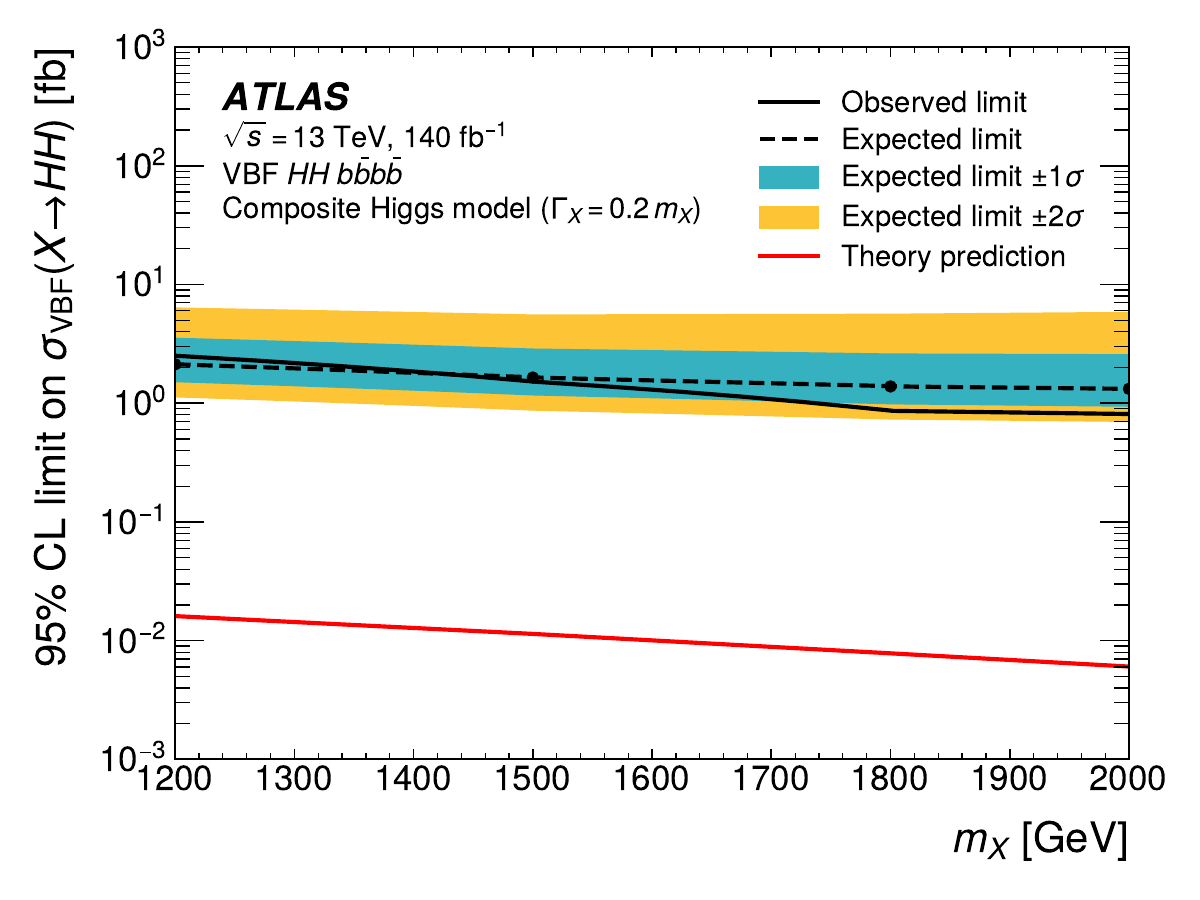}}
\caption{
Expected (dashed black lines) and observed (solid black lines) \SI{95}{\%} CL upper limits on the \xs of spin-0 heavy resonances with (a) narrow-width and (b) broad-width assumptions.
The SM \htobb branching ratio is assumed in both cases.
The $\pm 1 \sigma$ and $\pm 2 \sigma$ uncertainty ranges for the expected limits are shown as coloured bands.
The theoretical prediction for the Composite Higgs model calculated at leading-order~\cite{Contino_2011} under the $\Gamma_X/\mx = \SI{20}{\%}$ assumption is shown as the solid red line.
}
\label{fig:masslimit}
\end{figure}

Upper limits at the \SI{95}{\%} CL on the \xs for the narrow- and broad-width resonance assumptions are set in each available signal hypothesis using the asymptotic formula~\cite{Cowan:2010js} based on the \CLs method~\cite{Read:2002hq}.
\ifdoCONF
\else
The observed pBDT distributions of data, together with a signal + background fit, are shown in \cref{fig:pBDT} for certain narrow-width resonances.
\fi
The results are shown in \cref{fig:masslimit} for narrow- and broad-width resonances, which have masses in the range of $\SI{1}{\TeV} \le \mx \le \SI{5}{\TeV}$ and $\SI{1.2}{\TeV} \le \mx \le \SI{2}{\TeV}$, respectively.
The loss in sensitivity at high mass values is attributed to the smaller efficiency of the double $b$-tagging algorithm in the highly boosted regime.
The observed limits at \SI{1.6}{\TeV} (\SI{1.8}{\TeV}) and above drop for the narrow-width (broad-width) resonance assumption since no data are observed in the most signal-like bin of these high-mass pBDT distributions
\ifdoCONF
.
\else
, as shown in \cref{fig:pBDT}.
\fi
In the narrow-width assumption, the observed (expected) \SI{95}{\%} CL upper limit range spans from \narrowobslowmass~fb (\narrowexplowmass~fb) for $\mx = \SI{1}{\TeV}$ to \narrowobshighmass~fb (\narrowexphighmass~fb) for $\mx = \SI{5}{\TeV}$ and extends to \narrowobsmidmass~fb (\narrowexpmidmass~fb) for $\mx = \SI{1.8}{\TeV}$.
In the broad-width assumption, the observed (expected) \SI{95}{\%} CL upper limit range decreases from \broadobslowmass~fb (\broadexplowmass~fb) for $\mx = \SI{1.2}{\TeV}$ to \broadobshighmass~fb (\broadexphighmass~fb) for $\mx = \SI{2}{\TeV}$.
 
\ifdoCONF
\else
\begin{figure}[htb]
\centering
\subfloat[]{\includegraphics[width=0.5\textwidth]{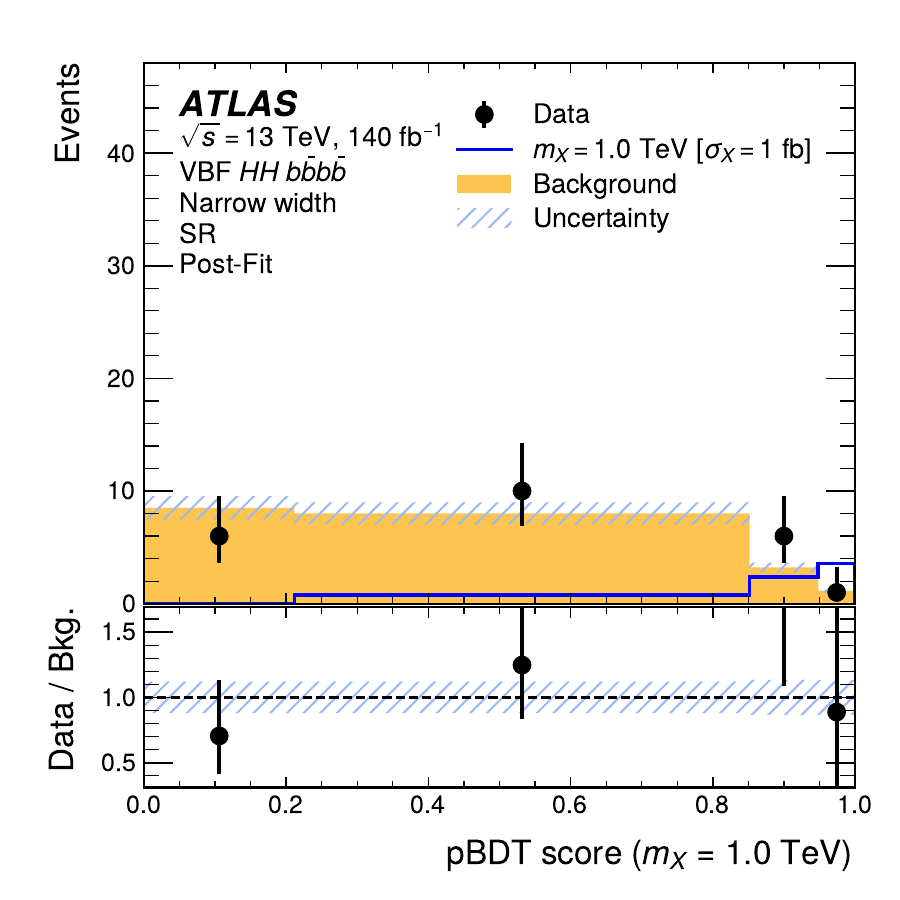}}
\subfloat[]{\includegraphics[width=0.5\textwidth]{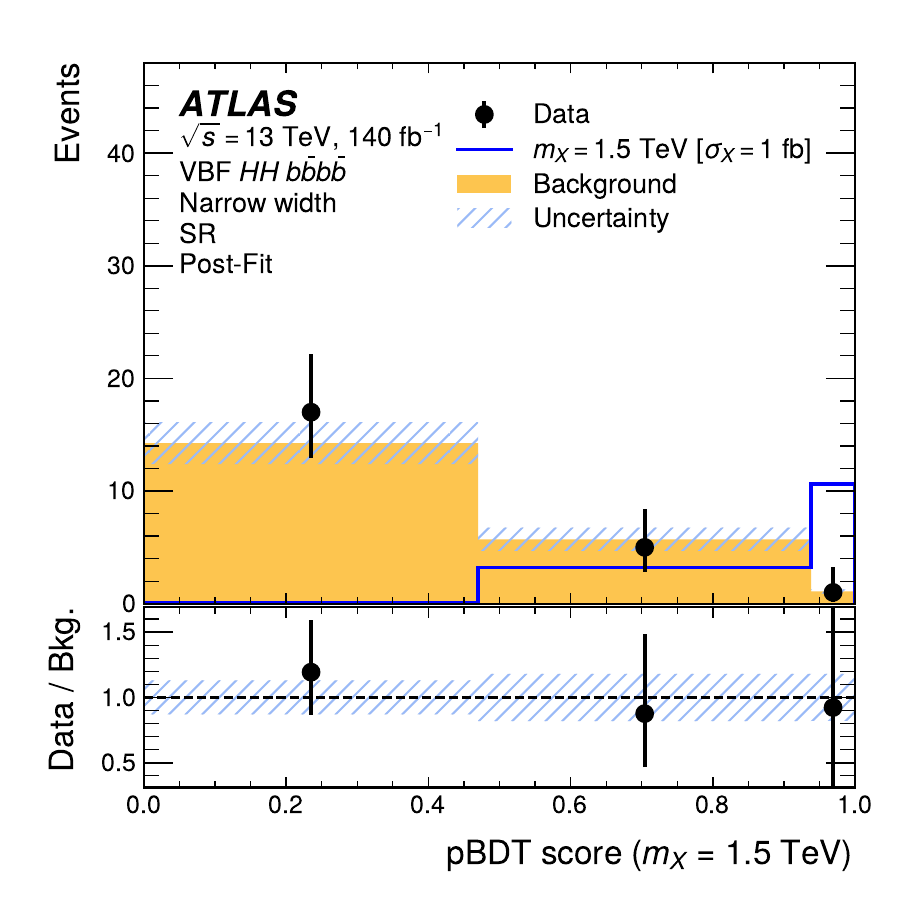}}\\
\subfloat[]{\includegraphics[width=0.5\textwidth]{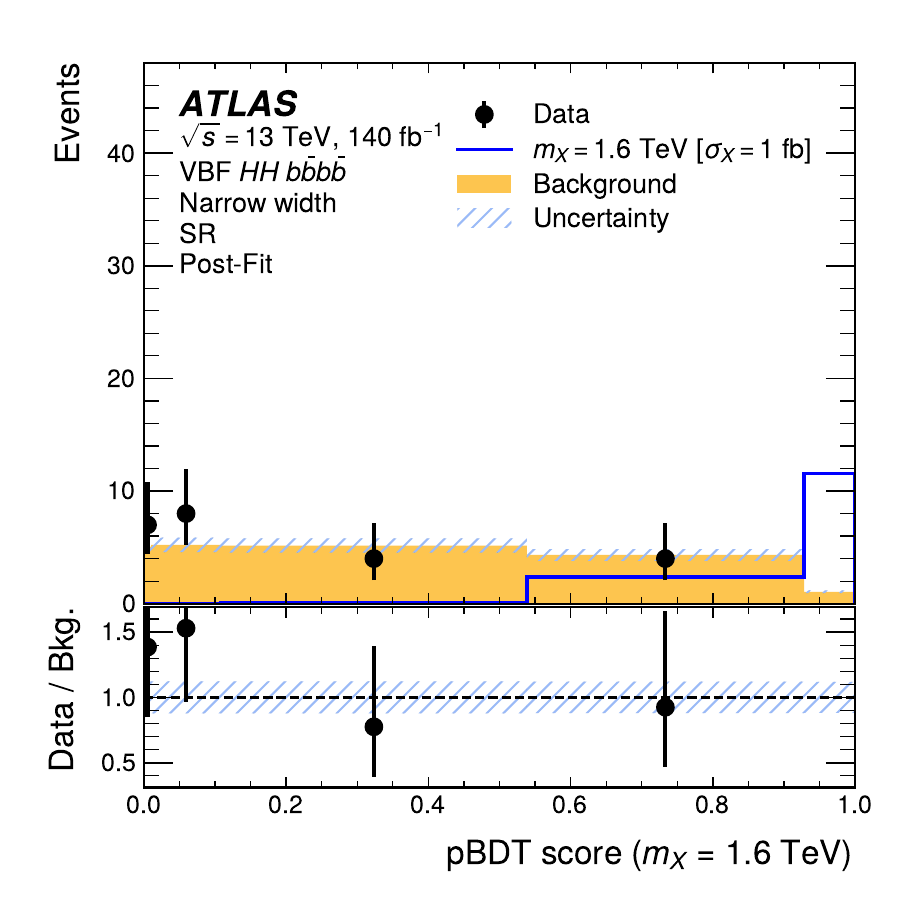}}
\subfloat[]{\includegraphics[width=0.5\textwidth]{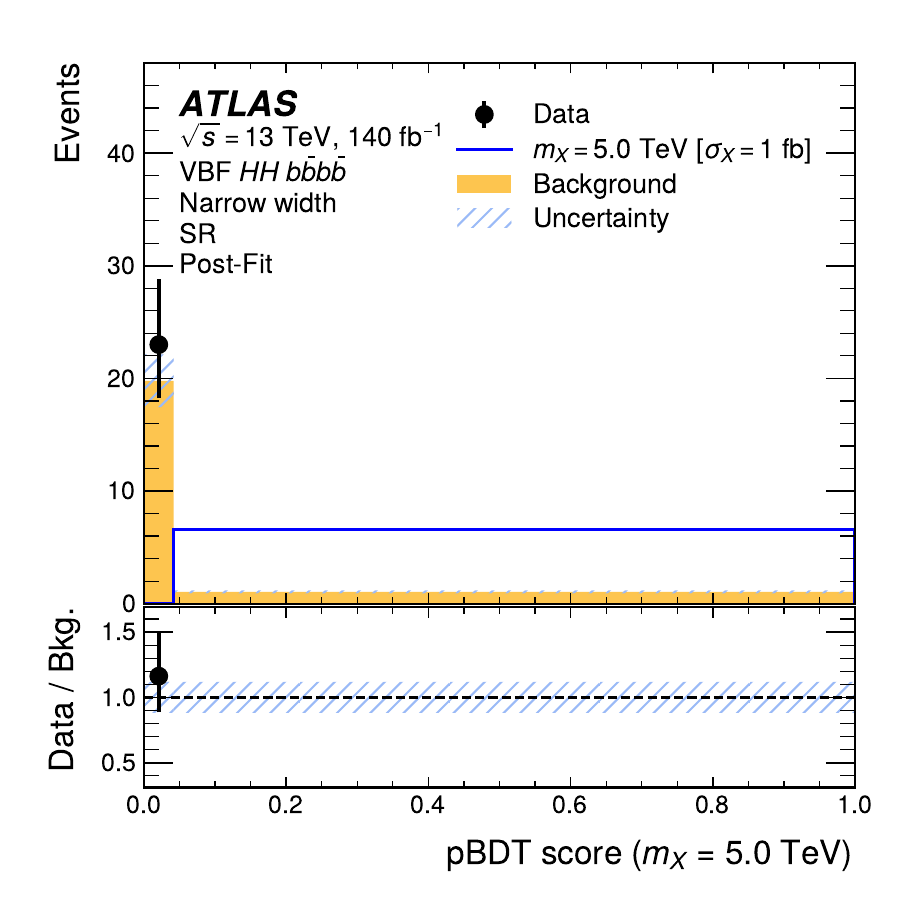}}\\
\caption{
The distributions of the mass-parameterised BDT score after a background-only fit to the data in the signal region.
The truth mass used as input to the mass-parameterised BDT corresponds to (a) $\mx = \SI{1.0}{\TeV}$, (b) $\mx = \SI{1.5}{\TeV}$, (c) $\mx = \SI{1.6}{\TeV}$, and (d) $\mx = \SI{5.0}{\TeV}$.
The narrow-width signal of the corresponding mass hypothesis, normalised to a \xs of \SI{1}{\fb}, is shown.
No events are observed in the rightmost bin for $\mx \ge \SI{1.6}{\TeV}$.
The binning procedure results in a very narrow first bin for (c).
The lower panel shows the ratio of data to the total prediction, with its uncertainty represented by the shaded band.
The error bars on the data points represent the statistical uncertainty.
}
\label{fig:pBDT}
\end{figure}
\fi
 
\FloatBarrier
\section{Conclusion}
\label{sec:conclusion}
A search for the production of Higgs boson pairs via VBF production in the four $b$-quark final state is presented.
The analysis is based on $\SI{140}{\ifb}$ of \pp collision data recorded with the ATLAS detector at the LHC, and focuses on the Lorentz-boosted regime where each Higgs boson is reconstructed as a large-$R$ jet.
This regime yields particular sensitivity to anomalous \kvv values that give rise to energetic Higgs bosons.
A machine learning-based double \Pqb-tagging technique is employed to enhance the analysis sensitivity, and boosted decision trees are used to discriminate signal from background.
The data are found to agree with the background-only hypothesis.
The observed (expected) constraints obtained are $\kvvobslow < \kvv < \kvvobshigh$ ($\kvvexplow < \kvv < \kvvexphigh$) at the \SI{95}{\%} CL\@.
The allowed range of \kvv values is reduced by a factor of two compared with previous ATLAS publications.
A value of $\kvv = 0$ is excluded with an observed (expected) significance of $\obssignificance$ ($\expsignificance$) standard deviations.
A search is also performed for the first time for a new heavy spin-0 resonance that would mediate VBF \HH production in a mass range between \SI{1}{\TeV} and \SI{5}{\TeV}.
No significant excess of events is observed and exclusion limits at the \SI{95}{\%} CL are set on the production \xs.


\FloatBarrier
 
\section*{Acknowledgements}


We thank CERN for the very successful operation of the LHC and its injectors, as well as the support staff at
CERN and at our institutions worldwide without whom ATLAS could not be operated efficiently.
 
The crucial computing support from all WLCG partners is acknowledged gratefully, in particular from CERN, the ATLAS Tier-1 facilities at TRIUMF/SFU (Canada), NDGF (Denmark, Norway, Sweden), CC-IN2P3 (France), KIT/GridKA (Germany), INFN-CNAF (Italy), NL-T1 (Netherlands), PIC (Spain), RAL (UK) and BNL (USA), the Tier-2 facilities worldwide and large non-WLCG resource providers. Major contributors of computing resources are listed in Ref.~\cite{ATL-SOFT-PUB-2023-001}.
 
We gratefully acknowledge the support of ANPCyT, Argentina; YerPhI, Armenia; ARC, Australia; BMWFW and FWF, Austria; ANAS, Azerbaijan; CNPq and FAPESP, Brazil; NSERC, NRC and CFI, Canada; CERN; ANID, Chile; CAS, MOST and NSFC, China; Minciencias, Colombia; MEYS CR, Czech Republic; DNRF and DNSRC, Denmark; IN2P3-CNRS and CEA-DRF/IRFU, France; SRNSFG, Georgia; BMBF, HGF and MPG, Germany; GSRI, Greece; RGC and Hong Kong SAR, China; ISF and Benoziyo Center, Israel; INFN, Italy; MEXT and JSPS, Japan; CNRST, Morocco; NWO, Netherlands; RCN, Norway; MNiSW, Poland; FCT, Portugal; MNE/IFA, Romania; MESTD, Serbia; MSSR, Slovakia; ARIS and MVZI, Slovenia; DSI/NRF, South Africa; MICIU/AEI, Spain; SRC and Wallenberg Foundation, Sweden; SERI, SNSF and Cantons of Bern and Geneva, Switzerland; NSTC, Taipei; TENMAK, T\"urkiye; STFC/UKRI, United Kingdom; DOE and NSF, United States of America.
 
Individual groups and members have received support from BCKDF, CANARIE, CRC and DRAC, Canada; CERN-CZ, PRIMUS 21/SCI/017, FORTE and PRIMUS, Czech Republic; COST, ERC, ERDF, Horizon 2020, ICSC-NextGenerationEU and Marie Sk{\l}odowska-Curie Actions, European Union; Investissements d'Avenir Labex, Investissements d'Avenir Idex and ANR, France; DFG and AvH Foundation, Germany; Herakleitos, Thales and Aristeia programmes co-financed by EU-ESF and the Greek NSRF, Greece; BSF-NSF and MINERVA, Israel; Norwegian Financial Mechanism 2014-2021, Norway; NCN and NAWA, Poland; La Caixa Banking Foundation, CERCA Programme Generalitat de Catalunya and PROMETEO and GenT Programmes Generalitat Valenciana, Spain; G\"{o}ran Gustafssons Stiftelse, Sweden; The Royal Society and Leverhulme Trust, United Kingdom.
 
In addition, individual members wish to acknowledge support from CERN: European Organization for Nuclear Research (CERN PJAS); Chile: Agencia Nacional de Investigaci\'on y Desarrollo (FONDECYT 1190886, FONDECYT 1210400, FONDECYT 1230812, FONDECYT 1230987); China: Chinese Ministry of Science and Technology (MOST-2023YFA1605700), National Natural Science Foundation of China (NSFC - 12175119, NSFC 12275265, NSFC-12075060); Czech Republic: Czech Science Foundation (GACR - 24-11373S), Ministry of Education Youth and Sports (FORTE CZ.02.01.01/00/22\_008/0004632), PRIMUS Research Programme (PRIMUS/21/SCI/017); EU: H2020 European Research Council (ERC - 101002463); European Union: European Research Council (ERC - 948254, ERC 101089007), Horizon 2020 Framework Programme (MUCCA - CHIST-ERA-19-XAI-00), European Union, Future Artificial Intelligence Research (FAIR-NextGenerationEU PE00000013), Italian Center for High Performance Computing, Big Data and Quantum Computing (ICSC, NextGenerationEU); France: Agence Nationale de la Recherche (ANR-20-CE31-0013, ANR-21-CE31-0013, ANR-21-CE31-0022, ANR-22-EDIR-0002), Investissements d'Avenir Labex (ANR-11-LABX-0012); Germany: Baden-Württemberg Stiftung (BW Stiftung-Postdoc Eliteprogramme), Deutsche Forschungsgemeinschaft (DFG - 469666862, DFG - CR 312/5-2); Italy: Istituto Nazionale di Fisica Nucleare (ICSC, NextGenerationEU), Ministero dell'Università e della Ricerca (PRIN - 20223N7F8K - PNRR M4.C2.1.1); Japan: Japan Society for the Promotion of Science (JSPS KAKENHI JP21H05085, JSPS KAKENHI JP22H01227, JSPS KAKENHI JP22H04944, JSPS KAKENHI JP22KK0227); Netherlands: Netherlands Organisation for Scientific Research (NWO Veni 2020 - VI.Veni.202.179); Norway: Research Council of Norway (RCN-314472); Poland: Ministry of Science and Higher Education (IDUB AGH, POB8, D4 no 9722), Polish National Agency for Academic Exchange (PPN/PPO/2020/1/00002/U/00001), Polish National Science Centre (NCN 2021/42/E/ST2/00350, NCN OPUS nr 2022/47/B/ST2/03059, NCN UMO-2019/34/E/ST2/00393, UMO-2020/37/B/ST2/01043, UMO-2021/40/C/ST2/00187, UMO-2022/47/O/ST2/00148, UMO-2023/49/B/ST2/04085); Slovenia: Slovenian Research Agency (ARIS grant J1-3010); Spain: Generalitat Valenciana (Artemisa, FEDER, IDIFEDER/2018/048), Ministry of Science and Innovation (MCIN \& NextGenEU PCI2022-135018-2, MICIN \& FEDER PID2021-125273NB, RYC2019-028510-I, RYC2020-030254-I, RYC2021-031273-I, RYC2022-038164-I), PROMETEO and GenT Programmes Generalitat Valenciana (CIDEGENT/2019/023, CIDEGENT/2019/027); Sweden: Swedish Research Council (Swedish Research Council 2023-04654, VR 2018-00482, VR 2022-03845, VR 2022-04683, VR 2023-03403, VR grant 2021-03651), Knut and Alice Wallenberg Foundation (KAW 2018.0157, KAW 2018.0458, KAW 2019.0447, KAW 2022.0358); Switzerland: Swiss National Science Foundation (SNSF - PCEFP2\_194658); United Kingdom: Leverhulme Trust (Leverhulme Trust RPG-2020-004), Royal Society (NIF-R1-231091); United States of America: U.S. Department of Energy (ECA DE-AC02-76SF00515), Neubauer Family Foundation.


\printbibliography
 
\clearpage
 
\begin{flushleft}
\hypersetup{urlcolor=black}
{\Large The ATLAS Collaboration}

\bigskip

\AtlasOrcid[0000-0002-6665-4934]{G.~Aad}$^\textrm{\scriptsize 104}$,
\AtlasOrcid[0000-0001-7616-1554]{E.~Aakvaag}$^\textrm{\scriptsize 17}$,
\AtlasOrcid[0000-0002-5888-2734]{B.~Abbott}$^\textrm{\scriptsize 123}$,
\AtlasOrcid[0000-0002-0287-5869]{S.~Abdelhameed}$^\textrm{\scriptsize 119a}$,
\AtlasOrcid[0000-0002-1002-1652]{K.~Abeling}$^\textrm{\scriptsize 56}$,
\AtlasOrcid[0000-0001-5763-2760]{N.J.~Abicht}$^\textrm{\scriptsize 50}$,
\AtlasOrcid[0000-0002-8496-9294]{S.H.~Abidi}$^\textrm{\scriptsize 30}$,
\AtlasOrcid[0009-0003-6578-220X]{M.~Aboelela}$^\textrm{\scriptsize 45}$,
\AtlasOrcid[0000-0002-9987-2292]{A.~Aboulhorma}$^\textrm{\scriptsize 36e}$,
\AtlasOrcid[0000-0001-5329-6640]{H.~Abramowicz}$^\textrm{\scriptsize 154}$
% ...
\bigskip
\\

% ...
$^{17}$Department for Physics and Technology, University of Bergen, Bergen; Norway.\\
% ...
$^{30}$Physics Department, Brookhaven National Laboratory, Upton NY; United States of America.\\
% ...
$^{36}$$^{(a)}$Facult\'e des Sciences Ain Chock, Universit\'e Hassan II de Casablanca;$^{(b)}$Facult\'{e} des Sciences, Universit\'{e} Ibn-Tofail, K\'{e}nitra;$^{(c)}$Facult\'e des Sciences Semlalia, Universit\'e Cadi Ayyad, LPHEA-Marrakech;$^{(d)}$LPMR, Facult\'e des Sciences, Universit\'e Mohamed Premier, Oujda;$^{(e)}$Facult\'e des sciences, Universit\'e Mohammed V, Rabat;$^{(f)}$Institute of Applied Physics, Mohammed VI Polytechnic University, Ben Guerir; Morocco.\\
% ...
$^{45}$Physics Department, Southern Methodist University, Dallas TX; United States of America.\\
% ...
$^{50}$Fakult\"{a}t Physik , Technische Universit{\"a}t Dortmund, Dortmund; Germany.\\
% ...
$^{56}$II. Physikalisches Institut, Georg-August-Universit\"{a}t G\"ottingen, G\"ottingen; Germany.\\
% ...
$^{104}$CPPM, Aix-Marseille Universit\'e, CNRS/IN2P3, Marseille; France.\\
% ...
$^{119}$$^{(a)}$New York University Abu Dhabi, Abu Dhabi;$^{(b)}$United Arab Emirates University, Al Ain; United Arab Emirates.\\
% ...
$^{123}$Homer L. Dodge Department of Physics and Astronomy, University of Oklahoma, Norman OK; United States of America.\\
% ...
$^{154}$Raymond and Beverly Sackler School of Physics and Astronomy, Tel Aviv University, Tel Aviv; Israel.\\
% ...
$^{a}$ Also Affiliated with an institute covered by a cooperation agreement with CERN.\\
$^{b}$ Also at An-Najah National University, Nablus; Palestine.\\
$^{c}$ Also at Borough of Manhattan Community College, City University of New York, New York NY; United States of America.\\
$^{d}$ Also at Center for Interdisciplinary Research and Innovation (CIRI-AUTH), Thessaloniki; Greece.\\
$^{e}$ Also at Centro Studi e Ricerche Enrico Fermi; Italy.\\
$^{f}$ Also at CERN, Geneva; Switzerland.\\
% ...

\end{flushleft}


\ifdoCONF
\else

\fi

\end{document}